\documentclass{mnras}
\usepackage{geometry}                		

\usepackage{graphicx}	
\usepackage{natbib}
\graphicspath{{/Users/andrew/Documents/TeXshop/contaminatedtempsDASH/Images/}}

\usepackage{newtxtext,newtxmath}
\usepackage[T1]{fontenc}

\DeclareRobustCommand{\VAN}[3]{#2}
\let\VANthebibliography\thebibliography
\def\thebibliography{\DeclareRobustCommand{\VAN}[3]{##3}\VANthebibliography}

\usepackage{xcolor}
\usepackage{makecell}
\usepackage{amsmath}
\usepackage{caption}
\usepackage{subcaption}
\usepackage{appendix}
\usepackage{comment}

\newcommand{\pkg}[1]{\textsc{\texttt{#1}}}
\newcommand{\snid}{\pkg{SNID}}
\newcommand{\ngsf}{\pkg{NGSF}}
\newcommand{\dash}{\pkg{DASH}}
\newcommand{\snnova}{\pkg{SUPERNNOVA}}
\newcommand{\orcid}[1]{\href{https://orcid.org/#1}{\includegraphics[height=11pt]{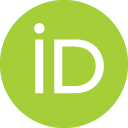}}}

\title{Testing and Combining Transient Spectral Classification Tools on 4MOST-like Blended Spectra}
\author[Milligan et al.]{
	A. Milligan$^{1}$
 \thanks{E-mail:a.milligan@lancaster.ac.uk}
 \orcid{ 0009-0006-6426-2431}
	I. Hook$^{1}$,
    C. Frohmaier $^{2}$ \orcid{0000-0001-9553-4723},
    M. Smith $^{1}$,
    G. Dimitriadis $^{1}$
    Y.-L. Kim$^{1, 14}$\orcid{0000-0002-1031-0796},
    K. Maguire$^{3}$,
    \newauthor
    A. M\"oller $^{4}$,
    M. Nicholl $^{5}$,
    S. J. Smartt$^{5, 6}$,
    J. Storm$^{7}$,
    M. Sullivan$^{2}$,
    E. Tempel$^{8}$,
    P. Wiseman $^{2}$ \orcid{0000-0002-3073-1512},
    L. P. Cassar\`a $^{9}$ \orcid{0000-0001-5760-089X},
    \newauthor
    R. Demarco$^{10}$,
    A. Fritz$^{11}$,
    J. Jiang$^{12},^{13}$
\\
$^{1}$Department of Physics, Lancaster University, Lancs LA1 4YB, UK \\
$^{2}$ School of Physics and Astronomy, University of Southampton, Southampton, SO17 1BJ, UK \\
$^{3}$ School of Physics, Trinity College Dublin, The University of Dublin, Dublin 2, Ireland \\
$^{4}$Centre for Astrophysics $\&$ Supercomputing, Swinburne University of Technology, Victoria 3122, Australia \\
$^{5}$Astrophysics Research Centre, School of Mathematics and Physics, Queens University Belfast, Belfast BT7 1NN, UK\\
$^{6}$Astrophysics Sub-department, Department of Physics, University of Oxford, Keble Road, Oxford OX1 3RH, UK \\
$^{7}$ Leibniz-Institut für Astrophysik Potsdam (AIP) An der Sternwarte 16, 14482 Potsdam \\
$^{8}$ Tartu Observatory, University of Tartu, Observatooriumi 1, Tõravere 61602, Estonia \\
$^{9}$ INAF-IASF Milano, via Alfonso Corti 12, 20133 Milano (Italy) \\
$^{10}$Institute of Astrophysics, Facultad de Ciencias Exactas, Universidad Andr\'es Bello, Sede Concepci\'on, Talcahuano, Chile \\
$^{11}$Kuffner Observatory, Johann-Staud-Strasse 10, 1160 Vienna, Austria \\
$^{12}$ Institute of Astronomy, University of Cambridge, Madingley Road, Cambridge CB3 0HA, UK \\
$^{13}$ Department of Physics, University of Warwick, Gibbet Hill Road, Coventry CV4 7AL, UK \\
$^{14}$ Department of Astronomy \& Center for Galaxy Evolution Research, Yonsei University, Seoul 03722, Republic of Korea
}
\date{Accepted XXX. Received YYY; in original form ZZZ}

\pubyear{2024}						

\begin{document}
\maketitle

\begin{abstract}
With the 4-meter Multi-Object Spectroscopic Telescope (4MOST) expected to provide an influx of transient spectra when it begins observations in early 2026 we consider the potential for real-time classification of these spectra. We investigate three extant spectroscopic transient classifiers: the Deep Automated Supernova and Host classifier (\dash{}), Next Generation SuperFit (\ngsf{}) and SuperNova IDentification (\snid{}), with a focus on comparing the completeness and purity of the transient samples they produce. We manually simulate fibre losses critical for accurately determining host-contamination and use the 4MOST Exposure Time Calculator to produce realistic, 4MOST-like, host-galaxy contaminated spectra. We investigate the three classifiers individually and in all possible combinations. We find that a combination of \dash{} and \ngsf{} can produce a SN Ia sample with a purity of 99.9\% while successfully classifying 70\% of SNe Ia. However, it struggles to classify non-SN Ia transients. We investigate photometric cuts to transient magnitude and the transient's fraction of total fibre flux, finding that both can be used to improve non-SN Ia transient classification completeness by 8--44\% with SNe Ibc benefitting the most and superluminous (SL) SNe the least. Finally, we present an example classification plan for live classification and the predicted purities and completeness across five transient classes: Ia, Ibc, II, SL and non-SN transients. We find that it is possible to classify 75\% of input spectra with >70\% purity in all classes except non-SN transients. Precise values can be varied using different classifiers and photometric cuts to suit the needs of a given study.

\end{abstract}

\begin{keywords}
transients: supernovae; techniques: spectroscopic; software: simulations; software: machine learning; instrumentation: spectrographs.
\end{keywords}

\section{Introduction} \label{intro}
Since the discovery of the accelerating expansion of the universe a quarter of a century ago \citep{1998AJ....116.1009R,1999ApJ...517..565P}, significant efforts have been made to investigate the enigmatic properties of dark energy. Many probes into the nature of dark energy exist, including weak lensing and Cosmic Microwave Background measurements \citep{2014A&A...571A...1P, 2000Natur.405..143W}. However, one of the most successful at providing strong constraints on cosmological models in the late-time universe is type Ia supernova (SN) cosmology. Understood to be the detonation of white dwarfs around the Chandrasekhar mass limit, SNe Ia detonate at predictable luminosities and as such act as standardisable candles that let us measure the distance to objects over large swathes of cosmic time.

The original discovery of accelerating expansion was performed with a sample of only 42 high-redshift SNe Ia \citep{1998AJ....116.1009R,1999ApJ...517..565P}. Since then, we have seen a two order of magnitude increase in the number of spectroscopically confirmed SNe Ia. For example, recently the Zwicky Transient Facility have produced their second data release sample (ZTF DR2) \citep{2025A&A...694A...1R} which contains 2,677 SN Ia with sufficiently high-quality light curves for use in cosmological fitting. Similarly, the recent Dark Energy Survey (DES) cosmology results \citep{2024ApJ...973L..14D} uses a sample of 1,635 SN Ia, derived from their full 5-year data release.

The earliest samples of transients were separated into two classes: SNe I and SNe II, based on the presence or absence of Hydrogen features in their spectra \citep{1937PASP...49..283P,1979sbaa.book.....L}. In the years since, these classes have been further subdivided and many new subclasses \citep{1997ARA&A..35..309F} and exotic variants have been discovered and suggested, alongside non-supernova transients like Tidal Disruption Events (TDEs) and Fast Blue Optical Transients (FBOTs) \citep{1975Natur.254..295H,2014ApJ...794...23D}.

Most optical transients are discovered in photometric surveys. As the number of transients has increased, it has become unfeasible to allocate time for spectroscopic follow-up on each transient individually. Recent photometric classifiers can perform high accuracy classification on transients beyond just classifying them as SN Ia or non-SN Ia \citep{2017ApJ...837L..28C,2019PASP..131k8002M,2019AJ....158..257B,2020MNRAS.491.4277M,2021AJ....162..275B,2023AJ....165...18P, 2024MNRAS.531.2474S, 2024A&A...689A.289C, 2025arXiv250101496S}. Additionally, it has been shown that they are capable of classifying transients based on incomplete light curves \citep{2020MNRAS.491.4277M,2022AJ....163...57Q,2023AAS...24110302G,2023ApJ...949..113G,2024ApJ...974..169D}. Recent photometric analyses have indicated that SN Ia samples obtained with photometric classifications produce contamination levels that either still allow for robust estimations of cosmological parameters or are even negligible compared to other sources of uncertainty, such as SN Ia astrophysics and how we model the correlation between SN Ia intrinsic properties and host-galaxy properties and how these instrinsic properties evolve with redshift. \citep{2018ApJ...857...51J, 2019ApJ...881...19J, 2024arXiv240102945V}.

While photometric classification is possible, it has several distinct disadvantages. The definitions of SN subclasses are based primarily by spectral features, so spectroscopic classification removes ambiguity, although there are also photometrically defined classifications. For example, SNe IIn are defined spectroscopically by narrow emission lines \citep{1990MNRAS.244..269S}, while SNe IIP are defined photometrically by a long `plateau' phase of constant brightness in their light-curve \citep{1997ARA&A..35..309F}. Further, when attempting to constrain cosmology, photometrically classified SN Ia samples often require the addition of spectroscopic information, such as spectroscopically determined host-galaxy redshifts. This is the case in \citet{2024arXiv240102945V}, where 1,635 photometrically classified SNe Ia are used for cosmology, the largest single-survey SN Ia sample. Additionally, \citet{2024arXiv240102945V} use a small sample of spectroscopically classified SNe Ia to constrain the cosmological fitting \citep[see also][]{2024ApJ...973L..14D}. Beyond this, to match the high purities of spectroscopically classified transient samples, photometric classification is usually performed in a binary scheme (SN Ia vs non-SN Ia) or with very broad transient classes \citep{2024arXiv240408798F}. 

We will, therefore, test the performance of spectroscopic classifiers. Visual classification is made difficult by the overlap of various transient subclasses in parameter space and ambiguity in subclass definitions. This, alongside the increasing number of transients being observed spectroscopically, means that it is increasingly required to automate the process of spectroscopic classification. We seek to investigate the potential to do this with regards to the upcoming 4MOST instrument.

The 4-metre Multi-Object Spectrograph Telescope \citep[4MOST]{2019Msngr.175....3D} is a high-multiplex, fibre-fed spectrographic survey facility in the final stages of assembly before commissioning. It is expected that it will begin taking data in early 2026. There are many varied surveys within the 4MOST consortium, but the survey concerned with transients is the Time Domain Extragalactic Survey (TiDES) \citep{2019Msngr.175...58S, 2025arXiv250116311F}.

With the upcoming Legacy Survey of Space and Time (LSST) being performed from the Vera C. Rubin Observatory, there will be unprecedented numbers of transients discovered photometrically \citep{2019ApJ...873..111I}. It is expected that any given pointing of 4MOST will contain a number of live photometric transients and the host galaxies of faded transients, which can then be followed-up with TiDES's allotted fibres. Over a period of 5 years, TiDES expects to observe 30,000 live transients and perform follow-up on some 200,000 host galaxies (these numbers are dependent on the survey schedules of LSST and 4MOST, both of which are still under development). This approach has already seen success in the Australian Dark Energy Survey (OzDES) performed using the AAOmega spectrograph on the Anglo-Australian Telescope \citep{2020MNRAS.496...19L,2004SPIE.5492..389S}.

Two of TiDES science goals are to provide live classification of transients accessible to the general scientific community and the classification of a large, pure, cosmological SN Ia sample. As we approach the start of the 4MOST survey in early 2026, uncertainty remains as to how the TiDES transient spectra will be classified and which existing spectroscopic classifiers, if any, are best suited to these two TiDES science goals.  Our hope is to provide clarity via the simulation of transient spectra that are as close to what will be observed as possible, including the fact that transient flux observed by a 4MOST fibre will be blended with the flux of its host galaxy. These realistic, blended, simulated 4MOST spectra will allow us to compare the output of various spectroscopic classifiers to known true classifications \citep[see also][which makes use of real spectra in its analysis]{2024arXiv241010963K}.
Furthermore, we can assess the dependence of classification performance on parameters such as the brightness of the SN and the fraction of host light contaminating the spectrum, and ultimately use this information to outline a plan for the classification of large numbers of TiDES spectra. 

There are two main types of automated, spectroscopic classifiers. First, there are template matching programs \citep[for example,][]{2009RAA.....9..341D, 2011asclsoft07001B, 2022TNSAN.191....1G}. These, in essence, compare an input spectrum to a bank of transients of known classification. However, there is significant variation in methodology. For example, \citet{2005ApJ...634.1190H} bin the input spectrum to match the templates and then calculate a \(\chi^2\) value, accounting for contaminant host flux. \citet{2011asclsoft07001B} instead cross-correlate input and template in redshift, and quantifies the best fitting template by the height of the cross-correlation peak.

More recent years have seen the rise of the second type: machine-learning methods \citep[for example,][]{2019ApJ...885...85M, 2020A&A...633A..88V,2021ApJ...917L...2F,2024arXiv241208601S, 2008A&A...488..383H}. In this case, a classifier is provided a training set of templates of known classification and redshift. The classifier "learns" the features present in various transient classifications and assigns them weights. The presence or not of these learned features is then used to determine a pseudo-probability of an input spectrum belonging to a given classification, which is then used to rank output classifications.

In this paper we investigate two template-matching classifiers and one machine-learning classifier. More information on the spectroscopic transient classifiers we investigate can be found in Sections \ref{DASH}, \ref{NGSF} and \ref{SNID}. These classifiers were chosen as they are publicly available, widely used and easily obtainable for current and upcoming surveys. Machine-learning algorithms are far faster to perform classifications once the lengthy training process is complete, but all classifiers as they are used in this work are expected to scale to TiDES.

Hence, this paper is organized as follows. First, in Section \ref{data}, we describe the simulations from which we draw our transient and host properties. Also in this Section we will discuss some transient templates used in simulating our blended spectra. In Section \ref{creating}, we will discuss the construction of blended host--transient spectra and the subsequent simulation of 4MOST observations using an Exposure Time Calculator (ETC). Then, in Section \ref{testing}, we investigate the capabilities of three individual spectroscopic transient classifiers. We go over their function and how they were tested. Their individual performances are presented in Sections \ref{ind results} and \ref{5 class}. We investigate the combination of classifiers in Section \ref{multiple}. We first show the results from a simple combination of classifiers and then potential photometric cuts for improving classification in Section \ref{ppcs}. Finally, in Section \ref{an example} we present a potential classification pipeline for live classification and SN Ia cosmology. Our conclusions are presented in Section \ref{conc}.

\section{Data} \label{data}

\subsection{Survey Simulations}

Our objective is to test spectroscopic transient classifiers such that we understand under what conditions they will succeed or fail in correctly determining the transient classes of 4MOST-like spectra. We must simulate a set of spectra that are a good approximation to the real ones observed by the instrument. The specific procedure for the creation of individual spectra is covered more in Section \ref{creating}, but we first discuss how we obtain a set a realistic properties for transients and their hosts. These properties can then be used to generate each spectrum, which in turn can be used to test each of the pre-existing transient classifiers. The results of these classifications can then be compared to the input spectrum's `true' properties as a means to quantify the success of a given classifier.
\\
We make use of two pre-existing, sequential simulations to produce a realistic sample of blended host-transient spectra. The first is a simulation of a population of transients and hosts performed in the SUpernova ANAlysis package \citep[SNANA]{2009PASP..121.1028K}. SNANA uses known intrinsic properties of various transient classes in combination with the survey strategy of the LSST survey to generate an LSST-specific transient population \citep{2025arXiv250116311F}. This simulation produces a population of transient and host objects. From them we obtain the intrinsic physical properties of host--transient systems. We obtain system redshift, host--transient separation, host \(r\)-band magnitude and transient template information. Throughout this paper magnitudes are calculated using the LSST \(r\)-band filter and are reported in the AB magnitude system \citep{1983ApJ...266..713O}. The process of creating simulated spectra is discussed in more detail in Section \ref{creating}.

\begin{figure*}
\centering
\begin{subfigure}{.49\textwidth}
  \centering
  \includegraphics[width=.9\linewidth]{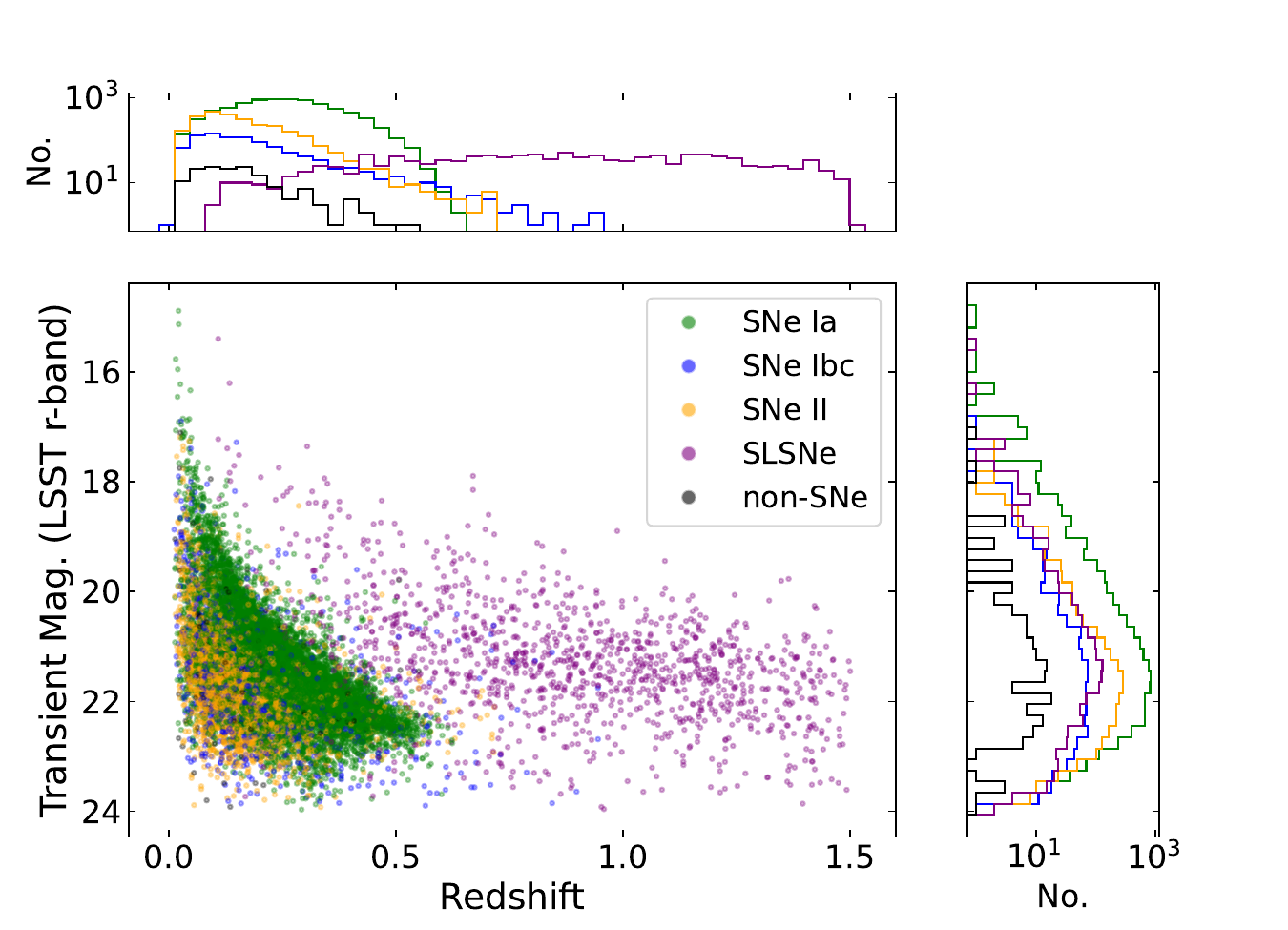}
  \caption{}
  \label{PP 1}
\end{subfigure}
\begin{subfigure}{.49\textwidth}
  \centering
  \includegraphics[width=.9\linewidth]{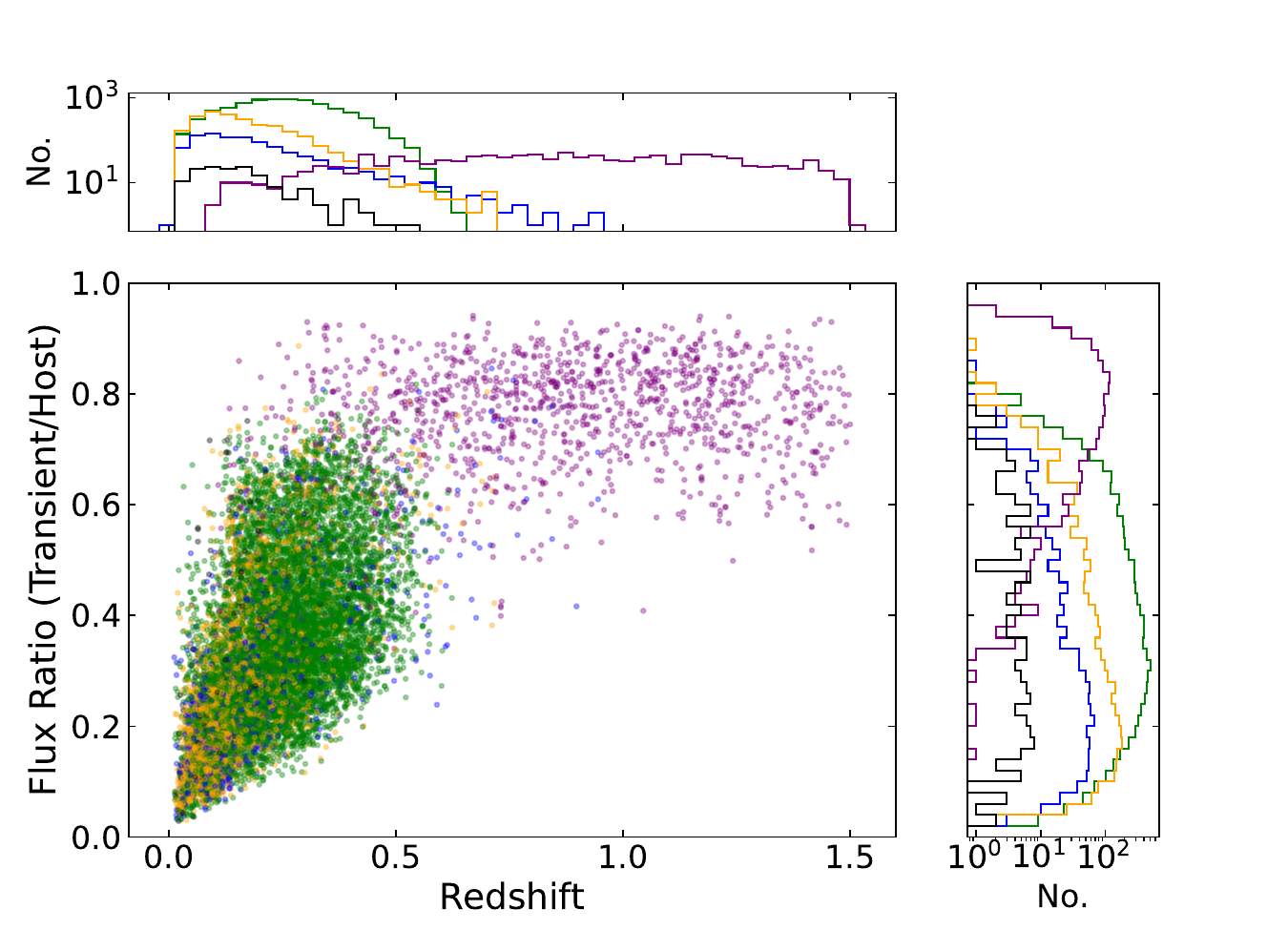}
  \caption{}
  \label{PP 2}
\end{subfigure}
\caption{a) Host galaxy redshift and corresponding transient magnitudes for observed objects in the SELFIE survey simulation. The values are obtained directly from the SNANA population simulation and can be considered the truth values for a given object. The y-axis on the attached histograms displays the total number of objects per bin with a logarithmic scale. b) As in (a) but with the fraction of fibre flux from the transient on the y-axis.}
\label{pop properties}
\end{figure*}

The second simulation is a simulation of the 4MOST survey operation of the full 5 years of observations of the southern sky. Observation targets are taken from the simulated survey input catalogs and their exposure times are computed using the 4MOST Exposure Time Calculator (ETC). The simulation is carried out with the 4MOST facility simulator (4FS) and makes use of the simulation code SELFIE. More detail about the SELFIE algorithm can be found in \citet{2020MNRAS.497.4626T, 2020A&A...635A.101T}. 

This simulation provides further observational properties for each transient. Most importantly, from it we receive a list of all of the transients that were observed. Generally, any transient that is both located within 4MOST's field of view during a visit, and is estimated to require less exposure time than is available during the full visit to meet the TiDES spectral success criterion (average \(SNR > 3\) in 15 \AA\ bins in the wavelength range of 4500-8000 \AA) will be observed. However, some are not observed due to the limited number of fibres and the demands of other subsurveys.

As the simulations have become more sophisticated, different versions of the input catalogue have been created. Each has had many different simulations of survey operations performed on it. We find that while the individual objects observed may change dramatically between simulations, the bulk properties of the observed transients are consistent. The specific simulation used has little effect on our final results.

The 4MOST observing schedule is currently expected to visit each sky position a small number of times during the 5-year survey. The survey footprint of 4MOST essentially covers the whole extragalactic sky in the Southern hemisphere. Each visit to a given position will consist of several exposures (most often 2 or 3) of approximately 20 minutes. The majority of transients (>93\%) are observed a single time over the course of the survey \citep{2025arXiv250116311F}.

The \(r\)-band magnitude, redshift and SN flux fraction distributions from the SNANA population simulation of the transients and their hosts from the SNANA population simulation are shown in Fig. \ref{pop properties}. The total number of objects in the sample is on the order of \(10^5\). We see that the sample is heavily biased to \(z < 0.6\) and in fact the more distant objects are all Superluminous Supernovae (SLSNe). We also see that, before any correction for fibre sizes, when observing extended objects (see Section \ref{Host FF}) there is a tendency for host galaxies to have brighter magnitudes than transients. 

\subsection{Simulated Spectra} \label{sim spec}

In addition to realistic physical and observational properties for use in creating simulated 4MOST-like spectra, we require a set of spectral templates of both transients and hosts. The transient templates are drawn from those used in the SNANA population simulations. The included SN classes are Ia, Ib, Ic, II, IIn, IIb and SLSNe. Most SNe Ia input templates are of the Ia-norm subclass, generated using the SALT2 model \citep{2007AA...466...11G}, although a small fraction are SNe Iax and SNe Ia 91bg-like \citep{2019PASP..131i4501K}. Additionally there are tidal disruption events (TDEs), and calcium-rich transient (CaRT) objects. These templates are spectral energy distributions (SEDs) intended to simulate realistic photometry. As a result, some of the spectra, especially SLSNe and non-SN transient, are highly smoothed and lacking in spectroscopic features. The full list of template sources is provided in Table \ref{sample makeup}. Examples of SEDs used in simulated blended spectra are shown in Appendix \ref{example spec}.

The galaxy templates from \citet{1996ApJ...467...38K} are assigned as hosts. The subclasses of galaxy available are elliptical, S0, Sa, Sb and Sc and a set of starburst templates with a variety of E(B-V) values (see \citet{1996ApJ...467...38K} for additional information). We scale our galaxy templates using the \(r\)-band host magnitudes from the simulation.

\begin{table}
\caption{The relative percentages of each transient class present in our full sample of blended spectra alongside the sources for the spectral templates. Templates can be found in the \texttt{SNANA} public data as part of PLASTICC \citep{2019PASP..131i4501K} and ELASTICC \citep{2023AAS...24111701N}.} 
\begin{center}
\begin{tabular}{c c c} 
Percentage & Class & Source\\
\hline
60.1\% & SNe Ia & \citet{2007AA...466...11G}, \citet{2018ApJ...867...23H} \\ 
0.9\% & 91bg-like & \citet{2019PASP..131i4501K} \\
1.1\% & SNe Iax & \citet{2019PASP..131i4501K} \\
1.9\% & SNe Ib & \citet{2019MNRAS.489.5802V} \\
1.4\% & SNe Ic & \citet{2019MNRAS.489.5802V} \\
13.5\% & SNe II & \citet{2019MNRAS.489.5802V} \\
6.5\% & SNe IIn & \citet{2019MNRAS.489.5802V} \\
4.0\% & SNe IIb & \citet{2019MNRAS.489.5802V} \\ 
9.4\% & SLSNe & \citet{2019PASP..131i4501K} \\
0.7\% & TDE & \citet{2019PASP..131i4501K}\\
0.4\% & CaRT & \citet{2019PASP..131i4501K}\\
\end{tabular}
\label{sample makeup}
\end{center}
\end{table}

For each transient we assign a host-galaxy morphology to match the probability distribution listed in \citet{2012A&A...544A..81H} in their Table 5. For Sd and Irregular galaxies for which we have no templates, we assign a random choice between Sb and Sc host spectra (the two most common host morphologies). In cases where \citet{2012A&A...544A..81H} lists the host as Morphology A/Morphology B, we choose randomly between A and B. We always assign SLSNe inputs an Sc-type host spectrum since research suggests that SLSNe are found in faint, blue, star-forming galaxies, often with extreme emission lines \citep{2015MNRAS.449..917L, 2011ApJ...727...15N}. TDEs and CaRTs  occupy such a small percentage of our transients, that we assign them a host type at random. However, we note that there is evidence that TDEs \citep{2024ApJ...960...69W} and CaRTs \citep{2022ApJ...927..199D} do show trends in their host galaxy morphologies, but including these in our simulations would have negligible impact in our results.

In order to estimate uncertainties in our results, we split the full sample of transients into samples of 1000 transients. This subsampling is performed randomly, but without resampling (i.e. no transient appears in more than one subsample). For a given parameter, results are obtained by reporting the mean value across all subsamples. The uncertainty on our results are reported in the form of the standard error of the mean.

\section{Creating Blended Spectra} \label{creating}

\subsection{The 4MOST Exposure Time Calculator}
The 4MOST ETC python code package\footnote{We use V2.3.1 of the python-based ETC: See \textsc{\texttt{qmostetc}} \href{https://escience.aip.de/readthedocs/OpSys/etc/master/index.html}{link to documentation}} allows one to simulate an observation by the 4MOST instrument. For every simulated observation we must assign a brightness within a specific filter or over a wavelength range. A variety of pre-existing instrument filters are provided.

The code produces  a `raw' or Level 0 (L0) output and a Level 1 (L1) output. Both are in the form of extracted 1D spectra (flux and wavelength for each pixel along the spectrum). The raw output features 4MOST's three spectrograph arms not yet combined and the object flux reported in ADUs. The L1 output is what we use. L1 spectra are generated by being passed through a simulation of the Quality Control 1 (QC1) pipeline and resemble the data products that will be produced by the real instrument. In L1 output, the ADUs of the raw output are converted to a flux observed at the telescope entrance using corrections for the wavelength dependence of the instrument's sensitivity. 

\begin{figure*}
\centering
\includegraphics[width=0.7\linewidth]{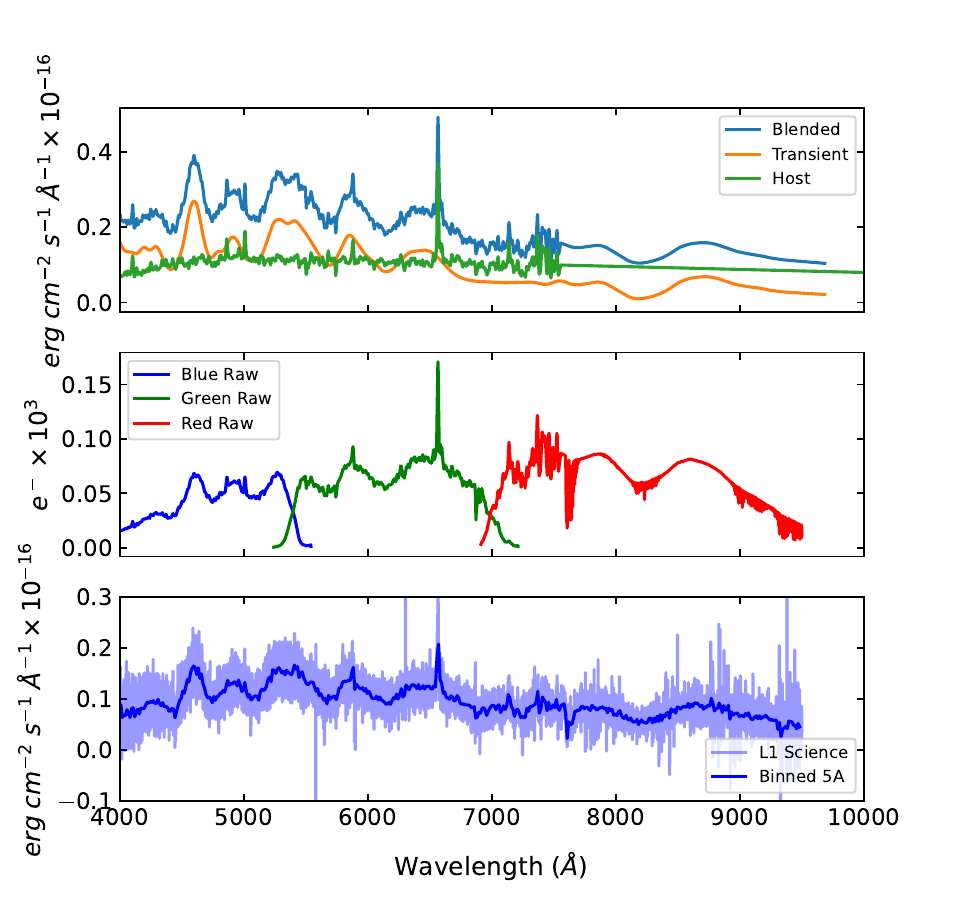}
\caption{The stages of simulating an observation with the 4MOST ETC code. In this example, a 21st magnitude SN Ia and a 21st magnitude Sc-type host spectra are added linearly. Top panel: template SN, host and combined spectra. All spectra are deredshifted. The flux is measured in units of \(\textup{erg} \; \textup{cm}^{-2} \; \textup{s}^{-1} \; \textup{\AA} ^{-1} \times 10^{-16}\). This is the input to the ETC. Middle panel : L0 output of the ETC, showing the extracted spectra from the three spectrograph arms. Flux is presented in units of \(e^- \times 10^3\). Lower panel: L1 output of the ETC in which the spectra from the three arms have been joined. The result is flux-calibrated and includes a realisation of the noise. This (unbinned) L1 spectrum is what we perform classification on.}
\label{blendedspec_process}
\end{figure*}

The simulation process is shown in Fig. \ref{blendedspec_process}. There are still telluric absorption bands present in the L1 output which are added as part of the ETC model. There are five main features with wavelength ranges of 6250 -- 6350, 6860 -- 6940, 7150 -- 7350, 7550 -- 7700 and 8100 -- 8400 \AA. These extra features could be misinterpreted by classifiers as being generated by the transient and lead to misclassifications. We account for this by creating a transmission spectrum for each observation. We do this on the assumption that real data will have these features corrected for using 4MOST observations of featureless calibration stars.

We consider the host and transient separately before adding them linearly to form the final spectrum that is input into the ETC for a simulated observation. The magnitudes of both objects are known from the population simulation, but to account for seeing conditions and a finite fibre size on extended galaxies we must adjust these magnitudes. The processes for doing so for SNe and galaxies are shown in detail in Sections \ref{SNe FF} and \ref{Host FF} respectively.

\subsection{Transient Fibre Flux} \label{SNe FF}
We assume the transient can be approximated as a point source and that the 4MOST fibre will be placed centrally on the transient. We simulate the fraction of transient flux through a 4MOST fibre using a grid of pixels with a central pixel containing the full transient flux. A Gaussian convolution is then applied to the pixel grid. The standard deviation, \(\sigma\), of the Gaussian convolution is determined from the Full-Width Half-Maximum (FWHM) of the seeing conditions using the expression \(\rm FWHM = 2\sqrt{2\ln 2} \sigma\). 

The SELFIE simulations do not record seeing conditions for each observation. For our purposes the seeing conditions are taken to always have a value of 0.8 arcseconds, this is similar to the average seeing conditions found at the Paranal Observatory where 4MOST will be located\footnote{From Paranal Observatory website, https://www.eso.org/gen-fac/pubs/astclim/paranal/seeing/?, accessed 23-January-2024}.

Once the Gaussian convolution has been applied, a fibre with a 4MOST fibre diameter of 1.45 arcseconds is imposed onto the pixel grid, centred on the SN location. The flux is then summed from the pixels with centres contained within the fibre radius. We find that using a finer pixel grid produces a more accurate value for fibre flux by reducing uncertainty around the fibre edge. This is particularly important in Section \ref{Host FF} where the scale of hosts being modelled varies and a balance must be found between accuracy and computation time. 

We are assuming a constant value for the seeing, coupled with a constant fibre size, so we see a constant fraction of transient flux down each fibre. The effect is that each transient appears 0.27 magnitudes fainter through the 4MOST fibre. This number does not require a simulation to be determined, as it determined from the integration of a 2D Gaussian out to some radius, but simulations are required for simulating extended hosts of varying size as discussed in Section \ref{Host FF}.

At seeing < 0.8 arcseconds the fraction of flux down the fibre from both transient and host is increased. Tests show that the increase is larger on average for transients (as they are point sources), so we would expect improved classification in this case. The reverse is true for seeing > 0.8 arcseconds and so we would expect worsened classification. Simulations indicated that increasing the seeing value to a uniform 1.2" had a small, negative effect on transient classification, but ultimately a realistic seeing distribution centred on 0.8" is expected to have minimal effect on the overall rates of successful transient classification.

\subsection{Host Fibre Flux} \label{Host FF}
The modelling of fibre flux from the transient's host galaxy, an extended object, is more complex. This method involves the dimensionless distance parameter (\(d_{DLR}\)), first used in \citet{2018PASP..130f4002S}, in service of assigning hosts to transients and based on similar methods developed in \citet{2006ApJ...648..868S}. The \(d_{DLR}\) is equal to the ratio of the directional light radius (DLR) of a galaxy and its observed separation from the transient. The DLR is the half-light radius of the galaxy in the direction of the transient. Minimising the \(d_{DLR}\) for galaxies in a crowded field indicates likely hosts for the transient. 

The population simulation we draw SNANA-produced physical properties from reports both the \(d_{DLR}\) and the host--transient separation. Since we are only concerned with the host's flux in the direction of the transient for the purposes of measuring the flux through a 4MOST fibre, we can consider all galaxies in the simulation to have circular half-light radii equal in radius to their DLRs. It should be noted that the position of the transient is entirely based on the light profile of the galaxy, so that transients are more likely to be placed in brighter regions of their hosts \citep{2021MNRAS.505.2819V}. 

We note that significant work has been performed investigating links between transients and their locations within their host galaxies \citep[see][ for example]{2016MNRAS.456.2848H, 2016MNRAS.459.3130A, 2018ApJ...855..107G}. However, since the population simulation preferentially places transients in brighter regions of their host, the resulting spectra may only be biased towards slightly higher levels of contamination from host flux. The effect on our results is negative, and is expected to be negligible.

We model the intensity of the galaxy to be a S\'ersic profile \citep{1963BAAA....6...41S} and use a S\'ersic index of 0.5 based on values reported in the simulations. While this may not be completely true to life, it represents the case with the most host flux in a blended spectrum and the hardest case to classify. Using a larger S\'ersic index causes the average host flux in the fibre to decrease leading to less host contamination. The S\'ersic profile is dependent on the value of the constant \(b_n\) which in turn is defined by the S\'ersic index. A number of approximations for the value exist such as \(b_n = 1.9992n - 0.3271\) for \(0.5 <= n <= 10\) from \citet{1989woga.conf.....C} and \(b_n = 2n - \frac{1}{3} + 0.009876n\) from the appendices of \citet{1997A&A...321..111P}. We will use the latter, although both produce very similar values for n = 0.5. 

The intensity profile, in terms of the S\'ersic index, \(n\), and \(b_n\), is often expressed as:

\begin{equation} \label{intensity profile}
I(R) = I_{e}\exp\{-b_n[(\frac{R}{R_e})^\frac{1}{n} - 1]\}
\end{equation}

where \(R_e\) is the effective or half-light radius that encircles half of the total emission of the profile. The effective intensity, \(I_e\), is the intensity at the effective radius.

To obtain the ratio of total galaxy flux to the flux transmitted through the fibre, we need to know the value of the total flux and the effective intensity. The total flux is obtained by integrating the intensity profile in equation \ref{intensity profile} which leads to the equation:

\begin{equation} \label{flux tot Ie}
F_T = 2.8941 \pi I_e R_e^2
\end{equation}

This gives us the total flux in terms of the effective intensity and the effective radius which is just the DLR \citep[for a more detailed derivation see][and references therein]{2005PASA...22..118G}. We can find the actual value of the total flux, and thus a value for the effective intensity, from the zero point magnitude of the AB magnitude system and the total magnitude of the galaxy, \(m_G\), using the equation: 

\begin{equation} \label{flux tot val}
 F_T = f_0 \times 10^{(m_{G} / -2.5)}
\end{equation}

Here \(f_0\) is the zero point flux of the AB magnitude system. The total host flux, \(F_T\), that appears in our equations, only functions as a scaling factor. We know the true value of \(m_G\) from the population simulation. By taking the ratio of total flux to flux in the 4MOST fibre, the value of the total flux cancels out and so it need not be calculated specifically. Once an arbitrary total flux is chosen we can calculate the effective intensity, \(I_e\), using equation \ref{flux tot Ie}. We can then use equation \ref{intensity profile} and equation \ref{flux tot Ie} to calculate the ratio between the total flux, the flux down the fibre and thus the host's magnitude as observed by 4MOST down its fibre.

We simulate a host's intensity profile by creating a pixel grid and use the S\'ersic profile to determine the average intensity at each pixel. Since we only care about the host's light profile in the direction of the transient, we model each host as a circle with a half-light radius equal to the DLR.

\begin{figure}
\centering
\includegraphics[width=\linewidth]{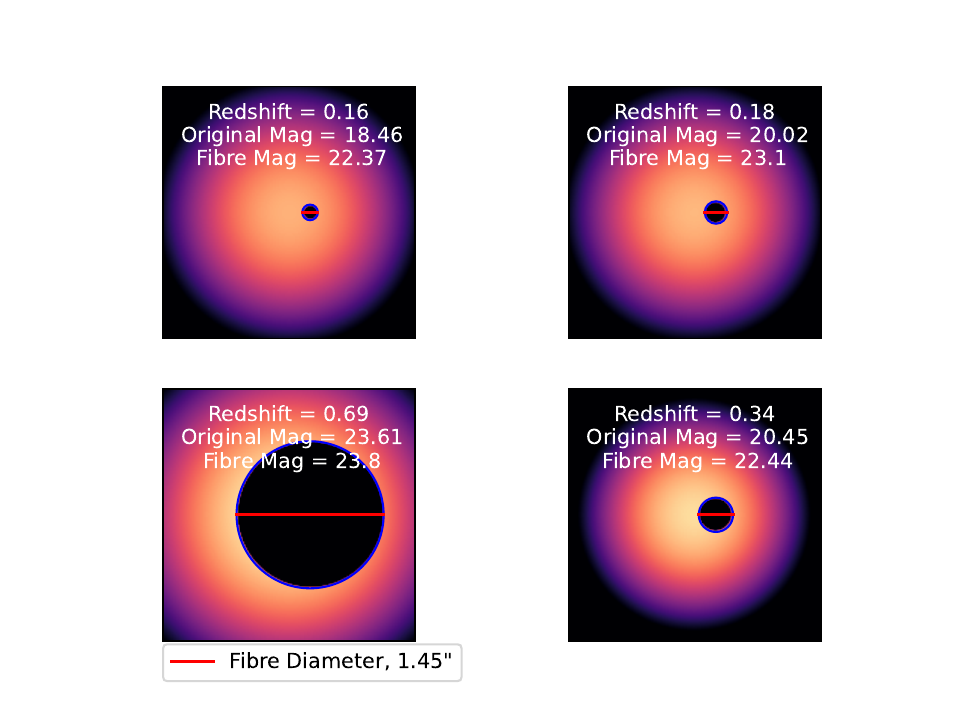}
\caption{The variation in the host flux through 4MOST fibres. Each panel presents the S\'ersic profile of an example host galaxy in our sample simulated on a pixel grid. Superimposed as a blue circle is the 4MOST fibre of diameter 1.45" (highlighted in red), centred on the transient location. The pixels that contribute to the flux seen by the fibre have their flux set to zero in these images, so that the lost flux can be seen. Redshifts, and host magnitude before and after accounting for fibre losses are provided.}
\label{extended}
\end{figure}

We then apply a Gaussian convolution to the pixel grid to account for atmospheric seeing. We use a 1,200 x 1,200 pixel grid with each pixel set to 1\% of the host--transient separation, a scale where the calculated flux fraction is invariant with small variations in pixel size. The method is identical to that described in Section \ref{SNe FF}. We centre the fibre on the transient location and calculate the fraction of flux in the fibre. Examples of this process are shown in Fig. \ref{extended}. We see much more significant flux loss than for the SNe.

\begin{figure}
    \centering    
    \includegraphics[width=\linewidth]{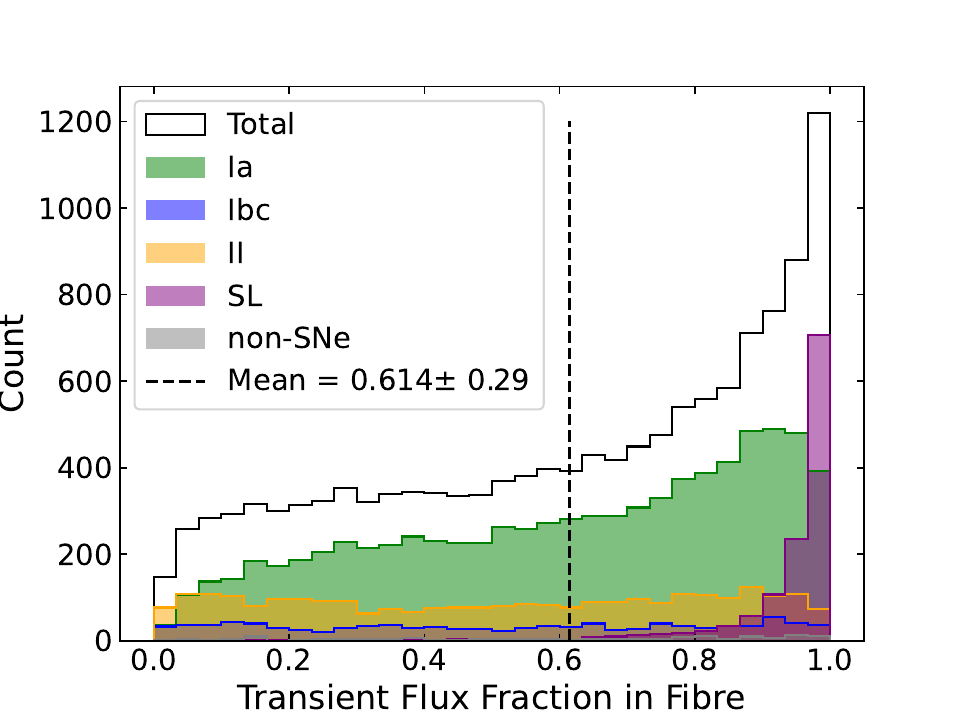}
    \caption{The distribution of transient flux fractions in the fibre. The mean value for all transients is highlighted with the dashed black line. As this accounts for fibre losses in the host galaxy, we see that over half of all of the spectra have more transient flux than host flux through the 4MOST fibre.}
    \label{transient ff hist}
\end{figure}

The 4MOST ETC cannot simultaneously account for both extended and point sources in a simulated observation. This is why we account for fibre losses and seeing effects ourselves, prior to passing the blended spectrum to the ETC. We provide the blended spectrum as being a flat illumination source with brightness measured in magnitudes per square arcsecond to prevent the ETC from reapplying any observational effects like seeing.

As stated in Section \ref{SNe FF}, the effect on the transient magnitude is fairly minimal. Most of the flux from the original point source still falls within the fibre that has a diameter of roughly \(2\sigma\) relative to the Gaussian convolution. For hosts, their distance, size and separation from their hosted transient result in significantly more variation in the fraction of the flux that is seen by the fibre (see Fig. \ref{extended}). This is a critical effect to model. By correcting the host magnitudes for fibre effects we see an average increase in the host magnitude of about 3.1 mag.

This leads to a reduction in host-galaxy flux contamination in the blended spectra. The distribution of transient fibre flux fractions shown in Fig. \ref{transient ff hist} demonstrates that we now have more than half of our spectra that are comprised of > 50\% transient flux over host. This has significance for spectroscopic classification as will be discussed in Section \ref{TFF cuts}.

\begin{figure*}
\centering
\includegraphics[width=0.7\linewidth]{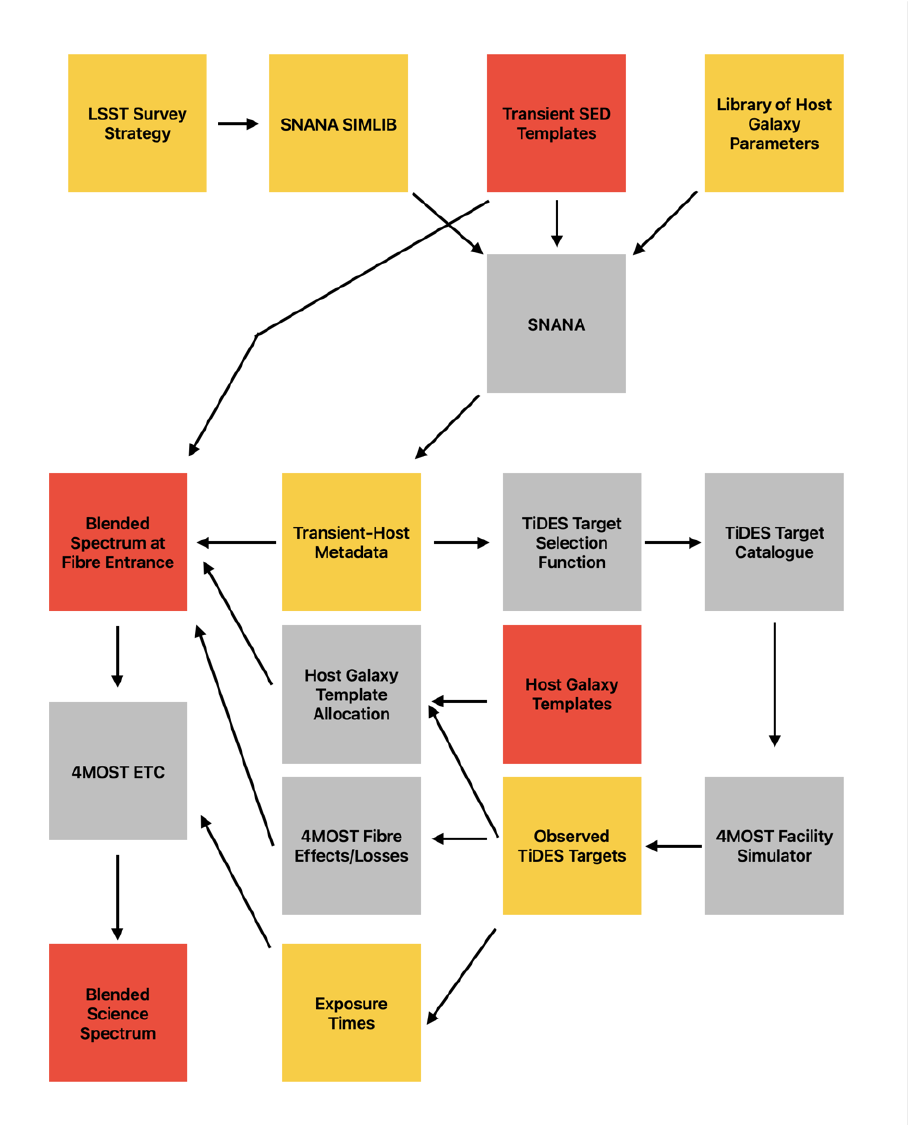}
\caption{Flowchart showing our simulation pipeline. Adapted from Figure 1 of \citet{2025arXiv250116311F}. Initially an LSST Operation Simulation (OpSim) is converted into a \texttt{SNANA} SIMLIB file. This, alongside a set of transient SEDs and a library of simulated host properties are used as inputs for a \texttt{SNANA} simulation that returns host-transient metadata and light--curves. These are input into the TiDES selection function as if operating in real--time. This produces a TiDES-specific target catalogue, for which the 4MOST facility simulator (4FS) generates fibre allocations and exposure times. This gives us a list of observed TiDES targets and their observational properties. Host galaxy templates are assigned to observed transients. The blended spectra have magnitudes and redshifts assigned from the \texttt{SNANA} metadata. Fibre losses are simulated to generate the spectrum at the 4MOST fibre entrance. This spectrum and its assigned exposure time from 4FS are input into the 4MOST ETC which adds realistic noise to the spectrum, producing our final blended science spectrum. Red boxes indicate templates or SEDs, yellow boxes indicate catalogue-level results and gray indicates a process or algorithm.}
\label{sim pipeline}
\end{figure*}

The full process used to create blended spectra as described across Sections \ref{data} and \ref{creating} is summarised in Fig. \ref{sim pipeline}.

\section{Individual Classifiers} \label{testing}

\subsection{Classifier Overviews} \label{overview}

\subsubsection{\dash{}} \label{DASH}

\dash{} is a deep convolutional neural network. \dash{} is trained on a set of templates and learns spectral features. Input spectra are broken down into individual features, compared to the features in the training set and then assigned a softmax pseudo-probability to each of its classification bins, named so due to the softmax regression model in the final layer of the deep learning model. The softmax probabilities only are only relative probabilities for one classification bin compared to the others \citep{2019ApJ...885...85M}. The highest pseudo-probabilities are then presented in the \dash{} GUI, and a combined softmax probability is produced by summing those of the best output bins until one is reached that either disagrees on transient class or is not in an adjacent phase bin. We discuss our method for converting the softmax probability for individual classification bins into probabilities for SN Ia, Ibc etc. in Section \ref{ind results}. The softmax probability of a classification bin is not necessarily a judgement on the quality of the classification. If every classification bin fits very poorly, then the best fit is not necessarily a good fit \citep{2019ApJ...885...85M}.

\dash{} also calculates an \(rlap\) cross--correlation value for each output classification bin as an additional flag for classification quality. The \(rlap\) parameter was  originally developed for another transient classifier that we investigate, \snid{}. However, we do not make use of it for \dash{}. 

\(rlap\) is the product of the correlation scale height ratio, \(r\), and \(lap\), an overlap parameter. \(r\) is defined as the ratio between the highest normalised cross-correlation peak, \(h\), and the root-mean-square (RMS) error of the anti-symmetric component of the cross-correlation product \(\sigma_{a}\):

\begin{equation}
r = \frac{h}{\sqrt{2}\sigma_{a}}
\end{equation}

\(lap\) is the overlap in \(ln(\lambda)\) space between the input and template spectra. A larger \(rlap\) value indicates more similarities between the input spectrum being classified and the template it is being compared to. Hence, larger \(rlap\) values indicate a better quality classification. The machine learning aspect of \dash{} returns the best-fitting classification bin. Then \(rlap\) values are calculated for each spectrum in \dash{}'s training sample in that classification bin. The highest \(rlap\) produced is returned to the user, with a warning if it less than five. Details on \dash{}'s template set can be found in \citet{2019ApJ...885...85M}. We do not make use of \(rlap\) in determining \dash{}'s classification results.

\dash{} has four modes of operation defined by its ability to fit or not fit transient host galaxies and its ability to use or not use known redshift values. We only make use of the known and unknown redshift modes. In the unknown redshift mode, the redshift is estimated by maximising \(rlap\) in redshift space. 

Host fitting leads to an increase in the number of output classification bins as each output now has a host class attached to each output. This increase in output bins leads diluted softmax percentages on outputs. 
However, we note that including a host-fitting step in the classification could remove degeneracy between transient class and redshift. Unfortunately, the host fitting mode doesn't function without redshifts provided. For this reason we do not investigate it.

There are some concerns that must be kept in mind if \dash{} is to be used as a mechanism to classify transients. For example, while \dash{} is user-friendly, fast-working and produces pure samples, it does so somewhat at the cost of user power. Compared to \snid{} or \ngsf{} \citep{2005ApJ...634.1190H} the user's options are fairly limited. There is no front-end mechanism to pass an error function for weighting the fit or removing wavelength ranges with known contaminant features. 

Additionally, and very importantly, the potential SN classes available for classification are somewhat limited. \dash{} can classify SNe Ia and common CC SNe like Ib/c, II, IIn and IIP. However, no other classes are included in its training sample and so other classes in the population simulations such as SLSNe, TDEs, and CaRT cannot be classified. They are either `other' results or contaminants. Some of these transient classes are fairly exotic and rare, but there are many SLSNe in the simulation and for \dash{} they can only act as a source of contaminant classifications.

\subsubsection{\ngsf{}} \label{NGSF}

Next-Generation SuperFit (\ngsf{}) is a template matching SN classifier. Written in python, it is based on the Superfit classification package written in IDL \citep{2005ApJ...634.1190H}. \ngsf{} requires a set of transient and host templates to compare to the spectrum being classified. We use the updated template set recommended in the source \footnote{From the \href{https://www.wiserep.org/content/wiserep-getting-started}{WISeREP} repository}. The input spectrum is sequentially compared to each of these templates while iterating through a variety of redshifts, reddening corrections and different levels of host contamination for a variety of morphologies. The redshift and reddening arrays that are checked are defined by the user. Each spectrum being fit must be compared to every template at every possible combination of reddening and redshift and for every host galaxy. As a result, the classification time required varies significantly with how fine the redshift sampling is \citep{2022TNSAN.191....1G}.

\ngsf{} returns its classification in the form of a \(\chi ^2\) value for each host, template, redshift, reddening combination. Input spectra are binned to match the templates and then a \(\chi ^2\) value is obtained using the equation (reproduced from \citet{2005ApJ...634.1190H}):

\begin{equation} \label{chi}
\chi ^2 = \sum \frac{[O(\lambda) - aT(\lambda; z)10^{cA_{\lambda}} - bG(\lambda; z)]^2}{\sigma(\lambda)^2}
\end{equation}

where \(O\) is the input spectrum, \(T\) is the transient template spectrum, \(G\) is the host galaxy template spectrum at a given redshift, \(z\), \(\sigma(\lambda)\) is the error on the input spectrum and \(A_{\lambda}\) is the reddening law. \(a\), \(b\) and \(c\) are constants that are varied during the classification process to check the template fit at varying reddening levels and at varying levels of host contamination. \ngsf{} uses the reddening law of \citet{1989ApJ...345..245C}. The templates with the lowest \(\chi^2_{red}\) is reported as the best template. As \ngsf{} also iterates through different levels of host contamination for each template it returns the estimated galaxy fraction of the best-fitting templates. Since our spectra have known SN and host magnitudes in the fibre, this has potential as another method to judge classification quality.

The throughput in the simulated 4MOST spectra drops below 70\(\%\) approximately below 4000 \AA\ and above 8000 \AA. We chose to limit the \ngsf{} template comparisons to this wavelength range. Since the ETC generates error spectra, we use these for calculating \(\chi^2\). In the case where the input spectrum has no attached error spectrum, \ngsf{} has several options for generating error spectra which can be used as weights to calculate a reasonable \(\chi^2\) for the input, although these are not the intended methods. It can determine a linear error spectrum or a Savitzky-Golay (SG) \citep{1964AnaCh..36.1627S} error spectrum.

The SG error spectrum is generated by smoothing the input spectrum with a SG filter and then subtracting the smoothed spectrum from the original to obtain residuals that are used to construct an error spectrum. The linear error spectrum is constructed using a linear fit to the binned input spectrum. In both cases this results in the smoothing of narrow features into noise, making both inferior to the use of an included error spectrum.

\ngsf{} has several distinct advantages over \dash{}, mainly in the form of user control. For example: the ability to set a redshift or reddening constant \textit{range} with specified values or the capacity to exclude noisy wavelength ranges. 

The final, and perhaps most considerable advantage, is \ngsf{} provides easy access to the set of templates it uses. This makes it very easy to update the templates manually to include more examples of existing subclasses or new subclasses altogether. Updates to either require no additional training time, which would be needed to change the templates used by \dash{}. \ngsf{}'s template set contains just over half as many transients as \dash{} and one third of the individual spectra, not including galaxy templates.

\subsubsection{\snid{}} \label{SNID}

\snid{} is an algorithm for determining the properties of a SN spectrum \citep{2007ApJ...666.1024B}. It makes use of cross-correlation techniques and the \(rlap\) quality parameter to find best-fitting redshifts, phases relative to maximum light and classes for input templates. \(rlap\) is discussed in more detail in Section \ref{DASH}.

We use templates collected from various samples by \citet{2022PASP..134b4505K}, where a more complete description can be found. Classifications were performed over the same 4000 - 8000\ \AA\ range as \ngsf{}. 

One advantage \snid{} has is the large variety of built-in transient classes and subclasses available for classification, as well as several morphologies of galaxy, AGN and a simple notSN classification amongst others that allow \snid{} to potentially identify non-transient spectra. \dash{} and \ngsf{} have no capacity to do this. \ngsf{} can easily have new templates added, but \dash{} would require computationally expensive retraining for the same effect. 

Further, addition of more subclasses is very simple. New templates can be added to the \snid{} repository provided they are in the correct format. Then the new classifications are added to a simple parameter file. In this paper we have 30 distinct classifications (a few SLSNe and non-SN classes were added to those that came built-in). However, \snid{} still seems to perform very poorly when classifying non-SN Ia spectra. This will be discussed further in Section \ref{5 class}.

One issue we encounter with \snid{} is that it occasionally performs a classification wherein none of its templates yield an \(rlap\) value greater than \(rlap_{min}\) and no output is produced. In this case we assign a best-fitting classification of `None' which is automatically considered an `other' classification.

\subsection{Classification Schema and Statistical Definitions} \label{Schema Defs}

With simulated transient spectra realistically blended with host galaxy flux now in hand, we can begin to test spectroscopic transient classifiers. We test the Deep Automated Supernova and Host classifier \cite[\dash{},][]{2019ApJ...885...85M}, Next Generation SuperFit \cite[\ngsf{},][]{2005ApJ...634.1190H} and SuperNova IDentification \cite[\snid{},][]{2011asclsoft07001B}. These classifiers are introduced in Sections \ref{DASH}, \ref{NGSF} and \ref{SNID}, respectively. Our objective is to compare the performance of each classifier on our simulated spectra.

\begin{table*}
\caption{The SN Ia and non-SN Ia transient subclasses for each classifier. The non-SN Ia transients subclasses included here match the various non-SN Ia input classes listed in Table \ref{sample makeup}. Any output classifications not included in this Table would be considered a misclassification if returned by a classifier.}
\begin{center}
\begin{tabular}{c c c c} 
 \hline
 Classifier & Binary Class & 5 Classes & Corresponding Outputs \\ 
 \hline
\dash{} & SNe Ia & SNe Ia & Ia-norm, Ia-91T, Ia-91bg\\
. & non-SN Ia & SNe Ibc & Ib-norm, Ib-pec, Ic-norm, Ic-broad \\
. &. & SNe II &Ib, IIP, II-pec, IIL, IIn \\
. & .& SLSNe & - \\
. &. & non-SN & - \\ 
. & . & Other & Ia-pec, Ia-csm, Ia-02cx \\
\ngsf{} & SNe Ia & SNe Ia &  \makecell{Ia-norm, Ia 91bg-like, Ia 91T-like, Ia 99aa-like} \\
. & non-SN Ia & SNe Ibc & Ibn, Ib, Ic, Ic-BL, Ic-pec, IIb \\
 .& .& SNe II & II, II-flash, IIn, IIb-flash \\
. &. & SLSNe & SLSN-II, SLSN-IIn, SLSN-I, SLSN-Ib, SLSN-IIb \\
. & .& non-SN & TDE H, TDE He, TDE H+He, FBOT, ILRT\\
. & . & Other &  \makecell{Ia 02es-like, Ia-02cx like, Ia-CSM-(ambigious), Ia-pec, Ia-CSM} \\  
 & & & \makecell{Ia-rapid, Ca-Ia, super-chandra, SN - Imposter, computed} \\
\snid{} & SNe Ia & SNe Ia & Ia, Ia-norm, Ia-91T ,Ia-91bg, Ia-99aa\\
. & non-SN Ia & SNe Ibc & Ib, Ib-pec, Ib-norm, Ic, Ic-norm, Ic-pec, Ic-broad, IIb \\
 .& .& SNe II & II, IIL, IIP, II-pec, IIn \\
 .& .& SLSNe & SLSN, SLSN-I, SLSN-Ic, SLSN-IIn\\
 .& .& non-SN & TDE, Ca-rich, ILRT\\
 . & . & Other &  \makecell{Ia-csm, Ia-pec, Ia-02cx,  NotSN, AGN, None} \\  
 & & & \makecell{LBV, M-star, QSO, C-star, LRN, Gal} \\
 \hline
\end{tabular}
\label{binary classes}
\end{center}
\end{table*}

The standards by which we will judge the performance of the classifiers are the purity and completeness of their classifications. Purity and completeness are, for a target transient class, defined as:

\begin{equation}
    \textup{Purity} = \frac{\textup{TP}}{\textup{TP} + \textup{FP}} \label{purity}
\end{equation}

\begin{equation}
    \textup{Completeness} = \frac{\textup{TP}}{\textup{TP} + \textup{FN}} \label{completeness}
\end{equation}

Here TP (true positive) are the number of spectra of the target class identified as such. FP (false positive) is the number of non-target class spectra misclassified as the target class.  FN (false negative) is the number of target class spectra misclassified out of the target class. TN (true negative) classifications are spectra correctly identified as not being in the target class.

Outside of binary classifications, for a given transient class, the completeness is the fraction of that class that are successfully identified as such. The purity is the fraction of output classifications of that class which are correct. Thus the rate of contamination in a transient class is 1 - purity for that class.

Throughout Sections \ref{testing} and \ref{multiple} we will, alongside completeness and purity, report the F-score (\(F_{\beta}\)) for each classifier \citep{van1977theoretical} as our figure--of--merit. \(F_{\beta}\) values range between 0 and 1 indicating a poor and a strong classifier, respectively. (\(F_{\beta}\)) is defined as:

\begin{equation}
    \textup{F}_{\beta} = \frac{(1 + \beta^2) \times \textup{Purity} \times \textup{Completeness}}{(\beta^2 \times \textup{Purity}) + \textup{Completeness}} \label{fbeta}
\end{equation}

\(\beta\) is a constant used to preferentially weight the \(F_{\beta}\) towards completeness or purity. The two main transient objectives of TiDES are the real--time classification of all transients from the TiDES-Live program and the eventual production of a SNe Ia sample for the purpose of fitting cosmology. The number of SNe we expect to obtain from 4MOST-TiDES is orders of magnitude larger than previous surveys such as Australian Dark Energy Survey \citep[OzDES,][]{2020MNRAS.496...19L} or the SuperNova Legacy Survey \citep[SNLS,][]{2006A&A...447...31A}. With the large number of spectroscopically observed  transients, we believe that purity is a more important factor than classification completeness. This is especially true for the SN Ia sample for cosmology, but even for real--time classification we choose to focus on pure samples.

With this in mind, we generally report the \(\beta\) = 0.5, \(F_{0.5}\), score as our Figure of Merit (FoM). This assigns greater weight to the classification purity over the \(F_1\)--score that weights both metrics equally. To account for multiple classes, each transient class has an individual \(F_{0.5}\) score calculated. Then the average value is obtained by taking the mean, weighted by each class's prevalence in the sample.

Additionally, in Section \ref{an example}, we will make use of the classification accuracy of our classifiers. This is particularly useful for comparison to photometric classifiers, which often use this parameter to quantify success. Accuracy is the fraction of classifications across all classes that are correct. In a binary schema it is defined as:

\begin{equation}
    \textup{Accuracy} = \frac{\textup{TP} + \textup{TN}}{\textup{TP + TN + FP + FN}}
\end{equation}

We do not aim for any particular purity threshold, but will add a 95\(\%\) purity line to relevant plots as an arbitrary point of comparison. This purity is similar to that found in SN Ia samples used in cosmology in the literature. For example, \citet{2005ApJ...634.1190H} reported an 8\(\%\) non-SN Ia contamination rate (92\(\%\) purity) in their final sample of SNe Ia, while \citet{2013ApJ...763...88C} reported a 3.9\(\%\) predicted contamination rate (96.1\(\%\) purity) that has an insignificant effect on their cosmological measurements. In \citet{2010A&A...523A...7G} purity ranges from 100\(\%\) to 90\(\%\) are found in various redshift bins up to \(z = 1\) and again, they report that the effect on cosmology is minimal compared to other sources of error.

Each classifier returns a list of output classification bins in descending order of the quality metric specific to that classifier. This is Softmax Probability (and \(rlap\)) for \dash{}, \(\chi^2\) for \ngsf{} and \(rlap\) for \snid{} as mentioned in Sections \ref{DASH}, \ref{NGSF} and \ref{SNID}, respectively. It is not clear if these quality metrics can be used in place of a probability or to what extent they can be compared. Additionally, as each classifier makes use of different templates either for training or matching, it is not necessarily reasonable to compare outputs from each classifier directly.

To determine the best output class for each classifier, we adapt the approach used in \citet{2023ApJ...942...29K}. A blended spectrum is input separately into each classifier. Then, for each classifier, the quality metric for each output classification is used to produce a probability that the input spectrum belongs to each of the output classes in the 5-class schema described in Section \ref{Schema Defs}.

For \dash{} this is a simple process as it already returns the Softmax pseudo-probability for each classification bin. We simply sum the softmax probabilities for the outputs corresponding to each of the five classes and normalise the resulting probabilities by the summed total of all Softmax probabilities.

For \ngsf{} we convert the returned \(\chi^2\) values into percentages by evaluating the cumulative density function at that particular \(\chi^2\). This is performed using the \texttt{scipy} python library. The resulting relative probabilities for each output are summed by class and normalised by dividing by the total probabilities for all outputs. When redshifts are provided the average number of reported outputs is 9.3. This jumps to over 50 when redshifts are not provided and often numbers of relatively spurious SLSN classifications can overweight that class as an output. To account for this we only look at up to the 10 best classifications when redshifts are not provided.

For \snid{} we are required to make a judgement call as the \(rlap\) quality metric it returns is less readily converted to a probability than those of \ngsf{} and \dash{}. In this case we obtain the value of \(r = rlap \times lap\) and convert it to a probability using the error function \(erf(r)\). For each class we sum the probabilities for each output in that class and then normalise these into probabilities by dividing by the sum of all output probabilities. We only consider such output classifications that meet the default \snid{} requirement of \(rlap_{min} = 5\). Because of this, all outputs return probabilities close to unity, meaning that we weight each output nearly equally.

Following these procedures provides us, for each spectrum for each classifier, the probability that the input is a SN Ia, Ibc, II, a SLSN, a non-SN transient or a non-transient (`other') spectrum. This standardisation of method allows for easy comparison of classification ability between the three classifiers.

We distinguish between SNe Ia that are `cosmologically useful’ and SNe Ia that are not. Ia-norm are counted as cosmologically useful, as are 91T--like SNe Ia. The latter are over--bright, hot SN Ia and are usually included in cosmological samples \citep{2025A&A...695A.140G}. SNe Ia 91bg-like standardisation for cosmology is debated \citep[see][and references therein]{2024MNRAS.530.4950G}. Here we consider them alongside Ia-norm inputs and output classifications. Any output that is not a SN Ia subclass is considered a non-Ia output.

To account for output classes for which we have no input spectra, we create the `other’ classification bin. This is a catch-all for automatic misclassifications from peculiar SN Ia subclasses (Ia-csm, Iax, etc.) or non-transient classes like `Gal’, `m-star’, `None’, etc. The list of `other’ classification outputs for each classifier are also included in Table \ref{binary classes}. For the purposes of calculating completeness, classifications that end up in the `other' class are considered FNs.

Some examples of successful and unsuccessful classifications are shown in Appendix \ref{App B}.

\subsection{Binary Classification Results} \label{ind results}

In this section we will be considering a binary classification. SNe will either be classified as a SN Ia or non-SN Ia. This is far fewer classes than each classifier has the potential to output, and we recognise that combining multiple output classes into a single, non-Ia class is not the same as requiring that a classifier chooses between two classes. We will also be tracking non-SN Ia transients that are misclassified as Ia contaminants. 

\begin{figure*}
\centering
\includegraphics[width=\linewidth]{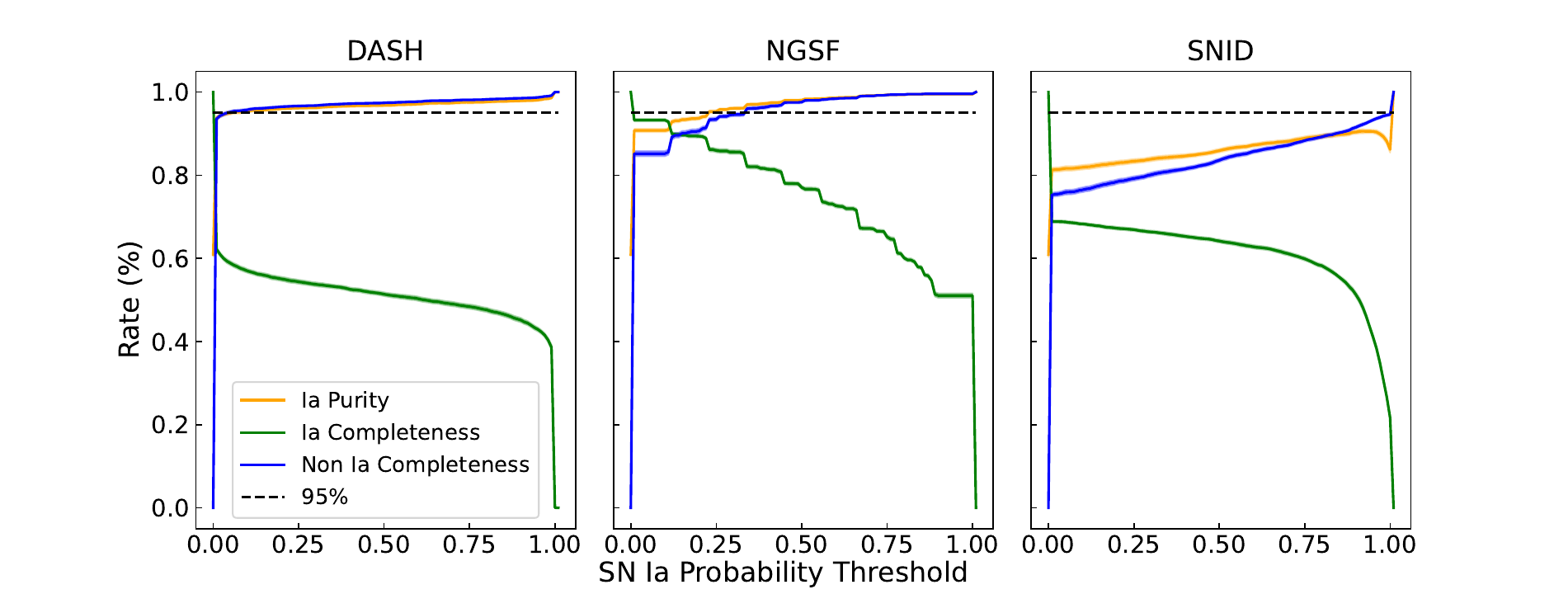}
\caption{SN Ia completeness (green), SN Ia purity (orange) and non-SN Ia completeness (blue) as a function of SN Ia probability threshold for each classifier. Input spectra are considered a SN Ia output if the returned SN Ia probability is greater than a given threshold, regardless of whether a different class is more probable. A rate of 95\% is marked by the dashed black line as an arbitrary point of comparison.}
\label{prob_limits}
\end{figure*}

Throughout this section, classification will be performed with known redshifts, simulating the case where a transient has a spectroscopic redshift determined from its host galaxy or its own emission features. In practice, this means that we provide the classifier with the true redshift from the simulation as a known redshift. Results for classification with unknown redshifts, or just photometric priors  are shown in Section \ref{5 class} and throughout Section \ref{multiple}.

We run the classifiers in non-interactive mode to mimic an automated classification plan for very large numbers of spectra. We note that this is not the way these classifiers were intended to run. Classifiers occasionally maximise their output metric with an incorrect classification, despite correct classifications being the second-- or third--best result. For example, this can occur where two output class are similarly favoured (say SN Ib and Ic) or where a completely spurious output classification is found due to redshift inaccuracy (a high--\(z\) SLSN classed as a low--\(z\) SNe Ia). By using all reported classifications from a classifier and converting to a probability for each of our output classes, we avoid this issue.

\begin{figure}
\centering
\includegraphics[width=\linewidth]{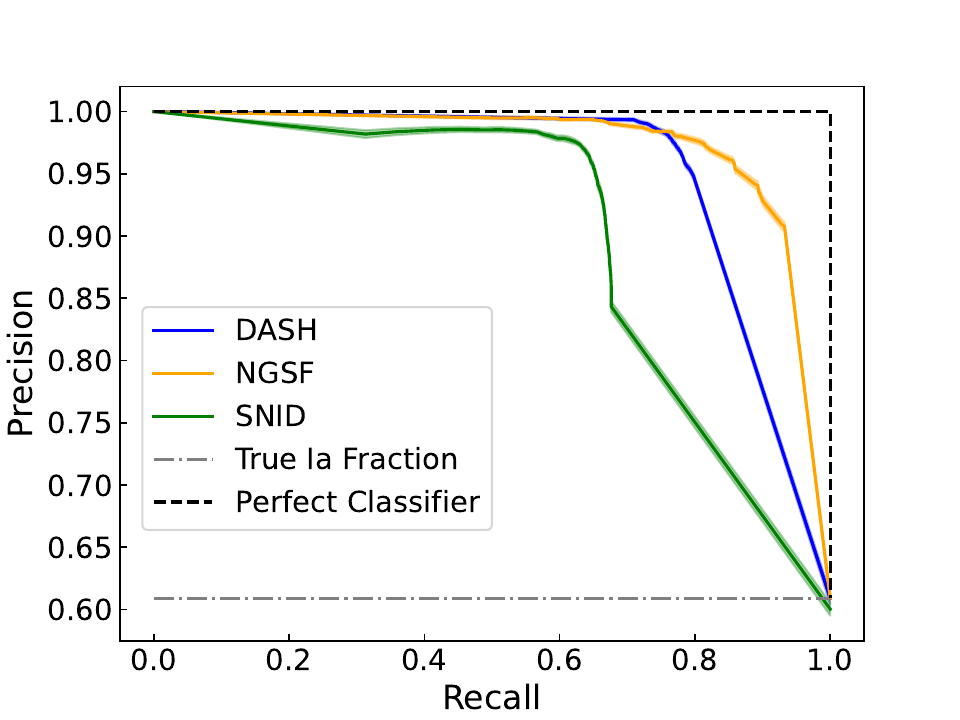}
\caption{Purity--completeness (precision--recall) curves for each of \dash{}, \ngsf{} and \snid{} in the case of binary SN Ia -- non-SN Ia classification. A theoretical, perfect, binary classifier is presented by the black dashed line. The closer a classifier's curve matches the perfect classifier, the better that classifier is performing. The grey dashed line indicates the fraction of input spectra that are SN Ia, which is the minimum possible purity obtained when the SN Ia probability threshold is set to zero.}
\label{pr curves}
\end{figure}

Our method of converting classifier outputs into probabilities returns the probability that a transient belongs to the SN Ia, Ibc, II, SL or non-SN transient classes defined in Table \ref{binary classes}. In this section we consider only the SN Ia probability and a binary SN Ia--non SN Ia classification schema. If the SN Ia probability exceeds an arbitrary threshold then that classifier will report it as a SN Ia, regardless of the probabilities of the other four classes. In Section \ref{5 class}, where we consider the full 5-class schema, we will swap to having the classifiers report each transient as whichever of the five classes has the greatest probability.

In Figure \ref{prob_limits} we investigate the SN Ia completeness, purity and non-SN Ia completeness for each classifier as a function of a SN Ia probability threshold. We can see that it is not immediately clear if a probability threshold should be applied for any of the classifiers. \dash{}'s SN Ia completeness, purity and non-SN Ia completeness remain almost constant for most SN Ia probability thresholds. Only at very low thresholds do we report purities under 95\% and only at very high thresholds do we see a large loss in SN Ia completeness. One could reasonably assign 0.5 as the required SN Ia probability to be considered a SN Ia.

Similarly \snid{} could reasonably have a SN Ia probability threshold set anywhere between 0.5 and 0.8. Below this we see significant losses to SN Ia purity and above this we see the same sudden loss in SN Ia completeness as displayed by \dash{}.

\ngsf{} is the only classifier to show a different trend. Here the SN Ia purity and non-SN Ia completeness quickly rise to unity. Meanwhile the SN Ia completeness starts at unity for no probability threshold, before steadily dropping as the threshold is made more stringent. A case could be made to perform \ngsf{} classification with a SN Ia probability threshold of anywhere from 10-25\%. Above this and the only change is a loss in SN Ia completeness.

In Fig. \ref{pr curves} we present purity--completeness (also known as precision--recall) curves for all three classifiers. A theoretically perfect classifier is shown as a point of comparison. A perfect classifier will return perfect purity at all levels of completeness as determined by varying the SN Ia probability threshold used to calculate each parameter. The only exception is the case where the threshold is set to zero. In this case the completeness is 100\% by definition, while the purity drops to match the fraction of the total input sample that are actually SN Ia, which is approximately 60\% in this case.

We can see from Fig. \ref{pr curves} that \ngsf{} performs closest to the theoretically perfect classifier. \ngsf{} is followed by \dash{} and \snid{} in that order. The uncertainty for each classifier, indicated by the transparent shaded regions around each curve, indicates an uncertainty on the order of ~0.5\%. This suggests that the classification results are stable across random samples of the full transient population. In other words, based on Figure \ref{pr curves}, we would expect  that \ngsf{} outperforms \dash{} and \snid{} across all of our subsamples under this binary classification schema. However, Figure \ref{pr curves} gives very little information about the non-Ia transients. For example, \ngsf{} could classify all SNe Ib as SLSNe and in this schema this would constitute perfect classification.

We do not report the numerical results for binary classification as the SN Ia classification is unchanged and allowing any non-Ia input to be `successfully' classified as any non-Ia output significantly inflates the non-SN Ia classification completeness and purity.

\subsection{Redshift Priors} \label{redshifts}

Using the SN Ia probability as a threshold gives a good indicator of the completeness and purities we can expect for each classifier and, also, allows use to construct purity--completeness curves that indicate that \ngsf{} is the best performing classifier in our binary schema. However, in this section we will proceed assuming that the output classification with the highest probability for each classifier is that classifier's output. This is partially to remove our need to assign arbitrary and distinct probability thresholds to each classifier and because it is the only method that is applicable for non-binary classification schemes. This avoids the situation where the SN Ia probability exceeds the threshold while being less than the probability that the transient belongs to a different class.

We test each classifier both with and without redshift priors. Using redshift priors means that for each input spectrum we provide the classifiers with the true transient redshift as found in the input population simulation. In the case of using unknown redshifts we give no redshift information to \dash{} and \snid{}. \ngsf{} is instructed to check redshifts between \(0 < z < 1.5\) with a sampling of \(\Delta z = 0.05\).

Perhaps one of the most likely scenarios during the operation of TiDES-4MOST is the case where we will not have a spectroscopic redshift, but will have a photometric redshift estimate. We would like to be able to investigate classifier performance in this scenario.

The minimum science requirement for LSST-DESC as reported in \cite{2018arXiv180901669T} is that the RMS scatter between photometric redshifts and true redshifts should not exceed 0.03\((1 + z)\). \citet{2018AJ....155....1G} and \citet{2023ApJ...944..212M} investigate LSST photometric redshifts instead assuming 0.02\((1+z)\) as the RMS error between photometric and spectroscopic redshifts. We will proceed using the 2\% uncertainty.

For \ngsf{} and \snid{} we are able to simulate the use of photometric redshift priors. We randomly generate a photometric redshift (\(z_{phot}\)) from a Gaussian distribution centred on the true redshift and with width equal to 2\% of \(1 + z\). Then we have each classifier attempt an `unknown' redshift classification over the truncated redshift range defined by a 2\% uncertainty in \(1+z_{phot}\). 

Unfortunately, \dash{} does not natively have the option to attempt classification over a custom redshift range. The only way for \dash{} to simulate photometric redshift priors is to have each classifier fit the randomly generated \(z_{phot}\) as a known redshift, which would prohibit a direct comparison to \ngsf{} and \snid{}. We found that this fitting of a `known', but slightly incorrect, redshift resulted in poorer performance than providing no redshift at all.

Because of this, we do not report on the classification potential of photometric redshifts throughout the paper. However, for completeness, we do report the results from \ngsf{} and \snid{} using them in the unknown redshift mode over a custom redshift range as described previously and making use of the 5-class classification schema as used in Section \ref{5 class}. These results are found in Table \ref{5class_comp} alongside the known and unknown redshift classification results. Additionally, when discussing combined classifiers in Section \ref{multiple}, we report the SN Ia completeness and purity for the combined \ngsf{}--\snid{} classifier using photo-\(z\) priors.

\subsection{5-Class Classification} \label{5 class}

In this section, we make use of a classification system that includes five transient classes: SNe Ia, SNe Ibc, SNe II, SLSNe and non-SN transients, following the work of \citet{2024arXiv241010963K}. The breakdown of classifier output subclasses that correspond to each of these inputs is indicated in Table \ref{binary classes}.

\begin{table*}
\caption{The completeness for classifying SNe Ia, SNe Ibc, SNe II, SLSNe, non-SN transients. Also presented are the SN Ia purity and the \(F_{0.5}\)--score for each classifier. The highest value in each column is highlighted in bold. Classification with photometric priors for \ngsf{} and \snid{} are provided alongside known and unknown redshift classification. \(F_{0.5}\)--score is calculated based on the average scores of all 5 transient classes reported, weighted by their population size.}
\begin{center}
\begin{tabular}{c c c c c c c c c}
 Classifier & Ia completeness & Ibc completeness & II completeness & SL completeness & non-SN completeness & Ia Purity & \(F_{0.5}\)--Score \\ 
 \hline
 \dash{}, known \textit{z} & 0.760 $\pm$ 0.004 & 0.68 $\pm$ 0.01 & 0.39 $\pm$ 0.01 & 0.0 $\pm$ 0.0 & 0.0 $\pm$ 0.0 & \textbf{0.981 $\pm$ 0.002} & 0.711 $\pm$ 0.003 \\
 \dash{}, unknown \textit{z} & 0.516 $\pm$ 0.004 & \textbf{0.69 $\pm$ 0.02} & 0.32 $\pm$ 0.01 & 0.0 $\pm$ 0.0 & 0.0 $\pm$ 0.0 & 0.968 $\pm$ 0.003 & 0.639 $\pm$ 0.003\\
 \dash{}, photo \textit{z} & -- & -- & -- & -- & -- & --\\
 \hline
 \ngsf{}, known \textit{z} & \textbf{0.798 $\pm$ 0.005} & 0.52 $\pm$ 0.02 & \textbf{0.753 $\pm$ 0.006} & \textbf{0.85 $\pm$ 0.01} & \textbf{0.05 $\pm$ 0.02} & 0.971 $\pm$ 0.002 & \textbf{0.814 $\pm$ 0.004}\\
 \ngsf{}, unknown \textit{z} & 0.560 $\pm$ 0.006 & 0.39 $\pm$ 0.02 & 0.35 $\pm$ 0.01 & 0.25 $\pm$ 0.01 & 0.02 $\pm$ 0.01 & 0.917 $\pm$ 0.003 & 0.627 $\pm$ 0.005\\
 \ngsf{}, photo-\textit{z} & 0.551 $\pm$ 0.006 & 0.48 $\pm$ 0.01 & 0.563 $\pm$ 0.008 & 0.85 $\pm$ 0.01 & 0.03 $\pm$ 0.01 & 0.935 $\pm$ 0.002 & 0.699 $\pm$ 0.003 \\
 \hline
 \snid{}, known \textit{z} & 0.661 $\pm$ 0.006 & 0.20 $\pm$ 0.01 & 0.174 $\pm$ 0.007 & 0.0 $\pm$ 0.0 & 0.0 $\pm$ 0.0 & 0.929 $\pm$ 0.004 & 0.649 $\pm$ 0.003\\
 \snid{}, unknown \textit{z} & 0.661 $\pm$ 0.006 & 0.15 $\pm$ 0.01 & 0.167 $\pm$ 0.006 & 0.0 $\pm$ 0.0 & 0.0 $\pm$ 0.0 & 0.835 $\pm$ 0.004 & 0.585 $\pm$ 0.005\\
 \snid{}, photo-\textit{z} & 0.644 $\pm$ 0.004 & 0.11 $\pm$ 0.01 & 0.083 $\pm$ 0.005 & 0.0 $\pm$ 0.0 & 0.0 $\pm$ 0.0 & 0.850 $\pm$ 0.005 & 0.552 $\pm$ 0.007\\
\end{tabular}
\label{5class_comp}
\end{center}
\end{table*}

Table \ref{5class_comp} shows the blended spectra being classified with the non-SN Ia transient output bin divided into SNe Ibc, SNe II, SLSNe and non-SN transients. 

The 5-class schema allows us to see finer detail about each classifier's ability to classify CC SNe and non-SN transients. This is particularly relevant for judging a classifier's ability to perform live TiDES classification across a range of different transient classes. $F_{0.5}$--scores reported throughout this section are the population size--weighted average of the $F_{0.5}$--scores of the five individual classes. The results from Table \ref{5class_comp} are presented as confusion matrices for the case with known redshifts in Fig. \ref{sing doub cm}(a).

As mentioned in Section \ref{redshifts} we take the output classification with the highest probability for each input spectrum as the output class, or best class. We impose no additional limit on the best class's probability beyond it being the highest probability. Across all classifiers we see small uncertainties (1-2\%) on purity and completeness, indicating that the classification rates are stable.

In every case the classifier's training sets are dominated by SNe Ia. This may lead to \dash{} over-weighting features learned from SNe Ia templates, resulting in an increased likelihood that a SN Ia classification bins will be amongst \dash{}'s top classification. Similarly, \snid{} and \ngsf{}, when the input does not match well with any of their templates, and lacking a redshift to help discount templates, are most likely to find SN Ia templates as the best matching templates as SNe Ia are the majority of their template banks. Across all classifiers, there is potential for SNe Ia to be the best matches in the absence of any good matches.

More detailed discussion on how input SN Ia spectra are being classified by \dash{}, \snid{} and \ngsf{} can be found in the appendix, in Fig. \ref{Ia fits}. Similarly, more detailed discussion on the origin of contaminant classifications for each classifier can be found in Fig. \ref{contaminant fits}.

\subsubsection{\dash{} Results} \label{dash res}

We see that our \dash{} results, both with and without redshift priors, have very impressive SN Ia purities well over 95\%. However, the SN Ia completeness, while fairly good with redshift priors, falls to just above 50\% without. This is the largest drop in performance upon the removal of redshift information, alongside \ngsf{}'s loss of SN Ia completeness. 

It becomes apparent that \dash{} is reasonably successful at classifying SNe Ibc when redshift priors are provided, but is far less successful at classifying type II SNe. Unlike what we see in its SN Ia completeness, when redshift priors are removed, there is not much change in performance for Type II SNe. The Ibc classification completeness actually improves slightly, while the Type II classification completeness decreases, but by far less than that of the SNe Ia. It cannot be stated strongly enough that \dash{} natively lacks all capacity to classify SLSNe and the various non-SN transients. Indeed, in Section \ref{multiple}, all combinations of classifiers that include \dash{} are incapable of successfully classifying any SLSNe or non-SN input spectra. 

Additionally, there is significant classification of input spectra into peculiar-Ia subclasses, often SN Ia-csm. This is particularly prevalent in transient spectra with Sc-type host galaxies, which make up a large fraction of our SN Ia hosts \citep{2012A&A...544A..81H}, likely due to emission lines present in the host template. The narrow emission lines from the host are misinterpreted as circum-stellar medium (CSM) interaction, leading to a Ia-csm classification.

Strangely, only \dash{}'s outputs exhibit this trend. Where 40\% of Sc-type hosts produce a Ia-csm classification in \dash{}, less than 1\% do in both \ngsf{} and \snid{}. Fortunately, this has no effect on classification purities in any class as SN Ia-csm is considered peculiar and outputs of Ia-csm are not included in final samples. However, it does have a significant effect on completeness.

\subsubsection{\ngsf{} Results}

\ngsf{} and \dash{} classify SNe Ia very similarly when redshift priors are provided. The difference in completeness for SN Ia (79.8.3\% v 76.0\%) is slightly in favour of \ngsf{}, the purity of the resulting SN Ia samples are almost identical, within 2 percentage points of each other. When removing redshift priors we see a loss of performance across Ia classification for both classifiers. The SN Ia classification completeness difference is similarly sized as in the case where redshifts are known, with \ngsf{} reporting 5\% higher completeness. However, while \dash{} reports only very slightly reduced (by less than a single percentage point) SN Ia purity, \ngsf{}'s corresponding rate drops by around 5 percentage points when redshift information is not provided.

In the 5-class scheme the finer non-SN Ia output classes leads to mixed classification results for \ngsf{}. The Ibc completeness is fair at just over 50\% with redshift priors. The non-SN transient completeness is very poor, well under 10\% with and without redshift priors (see Appendix \ref{example spec}), although \ngsf{} is the only classifier that gets any of these input spectra correct. \ngsf{} produces particularly impressive completeness in SN II and SLSN classifications when redshift priors are provided, but also reports drops in completeness of around 50 percentage points when redshift priors are not provided. This is still much better than \snid{}, which classifies no input SLSNe correctly, and \dash{} which, as mentioned previously, cannot classify them.

With redshift information, \ngsf{} is the strongest classifier in terms of classification completeness. Only \dash{} exceeds it in SNe Ibc completeness. Without redshifts, the balance between \ngsf{} and \dash{} is far closer due to \ngsf{}'s far larger loss of performance.

Indeed, when considering only the \(F_{0.5}\)-scores, \ngsf{} is now clearly the best performing classifier when redshifts are known. This is by a large margin, at least 0.1 larger than that of \dash{} or \snid{}. With unknown redshifts all three classifiers have \(F_{0.5}\) scores between 0.58 and 0.64. Here \dash{}'s score is heavily influenced by its superior SNe Ia purity, which is heavily weighted in our weighted $F_{0.5}$--score.

As would be expected, if a slightly incorrect photometric redshift (see Section \ref{redshifts}) with a small range of redshift values about it to consider is provided, performance improves compared to receiving no redshift at all. The $F_{0.5}$--score for \ngsf{} with photo-\(z\)s fall between that produced by known (spectroscopic) and unknown redshifts.

\subsubsection{\snid{} Results}

\snid{} has a much lower SN Ia completeness than \dash{} and \ngsf{} when given redshift priors, and with unknown redshifts we see a significant drop in performance in the SN Ia purity metric. However, without redshift priors we do see it outperform \dash{} and \ngsf{} in regards the SN Ia completeness. In fact, its SN Ia completeness is nearly invariant under a lack of redshift information. However, while the SN Ia completeness is maintained, this must be balanced against the significant drop in SN Ia purity, which leads \snid{} to a poorer $F_{0.5}$--score than \dash{} or \ngsf{} without redshift information.

\snid{} produces poor classification completenesses in all non-SN Ia transient subclasses in the 5-class schema. With or without redshift information it only achieves SN Ibc and II completenesses between 10\% and 20\%. Like \dash{} it classifies no SLSN or non-SN transient correctly, but while \dash{} is incapable of outputting such classifications, \snid{} instead fails to do so. A large number of our blended spectra are classified as `Gal' (a galaxy template) by \snid{}, leading to an `other' output. It appears that galaxy contamination may be a limiting factor. Indeed, \ngsf{} is trained to classify host and transient simultaneously which may explain its superior performance.

When photometric classification is possible, the results are the opposite of that seen with \ngsf{}. For all transient classes with classification completeness greater than zero without redshift information, the completeness is lower with photometric priors. \snid{}'s SN Ia purity does improve with photometric redshifts relative to a lack of redshift information, but the final $F_{0.5}$--score is still lower. SLSNe are well classified by \ngsf{}, as photo-\(z\)s force the classification into the superluminous regime, yet this doesn't appear to occur in \snid{}.

It should be noted that \snid{} was intended to have significant human oversight in classification, so relatively poor results under complete automation are not unexpected. Additionally, while \snid{}'s $F_{0.5}$--score is lower than the other two classifiers, its $F_1$ or $F_2$ scores are not. As \snid{} maintains SN Ia completeness when redshifts are unknown, and so \(F_{\beta}\)--scores that are weighted to more heavily favour completeness (\(\beta > 1\)) lead to \snid{} matching \ngsf{}'s performance and exceeding \dash{}'s when redshifts are unknown.

\section{Using Multiple Classifiers at Once} \label{multiple}

\begin{figure*}
\centering
\begin{subfigure}{0.9\textwidth}
  \includegraphics[width=.9\linewidth]{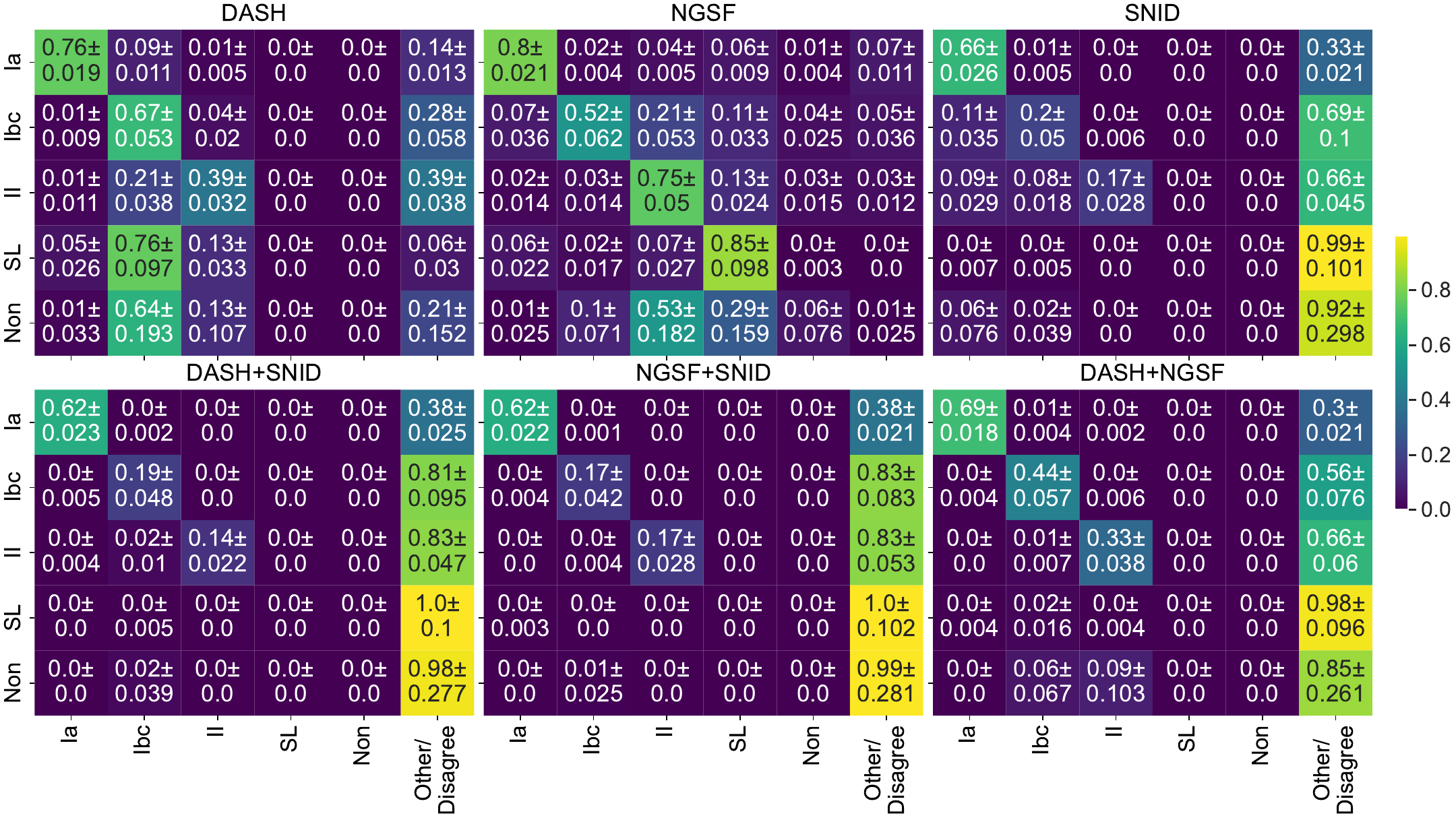}
  \caption{Normalised by Row}
  \label{sing class}
\end{subfigure}
\begin{subfigure}{.9\textwidth}
  \includegraphics[width=.9\linewidth]{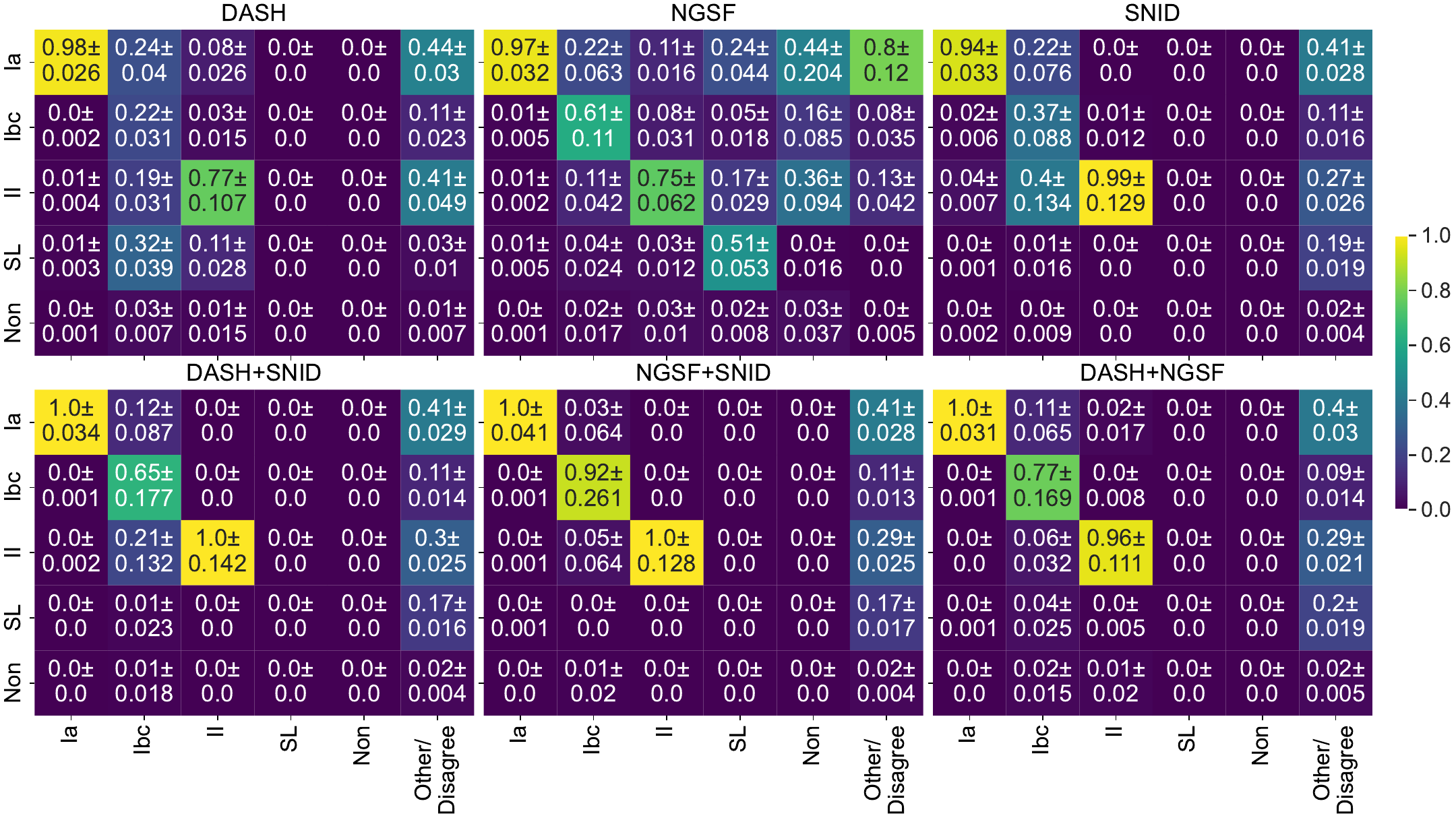}
  \caption{Normalised by Column}
  \label{doub class}
\end{subfigure}
\caption{Confusion matrices showing the results for the three individual classifiers and all three combinations of two of the three classifiers working simultaneously. Confusion matrices are normalised by a) row, indicating completeness in each class and b) column, indicating the purity of each class. The `other' output classification is reserved for output classifications with no corresponding input class and, in the case of the combined classifiers, an input spectrum that causes the two classifiers to disagree on the output class. Classification was performed with redshift priors provided in all cases. High completeness and purity samples would be indicated by high concentration along the matrix diagonal. Horizontal scatter indicates loss of completeness, vertical scatter indicates loss of purity.}
\label{sing doub cm}
\end{figure*}

For both live classification of transients and when creating SN Ia samples for cosmology, it is critical to limit contamination in the output sample. For live classification, this is important for all SN classes. For cosmology, it only matters that the SN Ia sample is of high purity, even to the detriment of the SN Ia completeness. This is particularly true given the very large number of transients that 4MOST is expected to observe.  Table \ref{5class_comp} shows that individual classifiers struggle to limit contamination in the output SN Ia sample and are poor classifiers of even broad non-Ia SN classes. The obvious question is: what is the result of combining the classifications from different classifiers for each transient?

We first investigate the effect of classifying spectra with all combinations of two out of the three classifiers. In these cases, if both classifiers are not in agreement on the output classification, then the result defaults to an `other' output regardless of the quality of either classification. Any output classifications from individual classifiers that do not match any of our potential output classes (Ia-pec, non-transients, etc.) are also discarded as `other' outputs.

Fig. \ref{sing doub cm} shows that when using known redshifts, requiring two classifiers to agree has the effect of reducing the overall completeness for all five original output classes and a large increase in the number of `other' outputs compared to the individual classifier results. However, we also see a large increase in the purity of SNe Ia, SNe II and, to a lesser extent, SNe Ibc. This can be seen by high concentrations along the confusion matrix diagonals.

The extreme case for a combined classifier is to use all of \dash{}, \ngsf{} and \snid{} simultaneously. The results for SNe Ia are shown in Table \ref{combined puritys}. With the combination of all three classifiers, we now classify around 60\(\%\) of all SNe Ia when redshifts priors are provided, but get very few successful classifications for any other input class. The sample of classified SNe Ia produced by this combined classification is completely pure.

Without redshifts we report reduced success. While SN Ia purities remain very high, the non-SN Ia completenesses remain around 10\% or less and the SN Ia completeness is nearly halved to 33\%. This is very low compared to other combined and individual classifiers. It remains to be determined where exactly the optimum balance lies between pure SN Ia samples and large SN Ia samples for the purposes of cosmology. Regardless, combined classification has the promising ability to improve SNe Ia, II and, to a lesser extent, SNe Ibc purity.

Using all three classifiers, 87\% of SNe II are misclassified as `other' or SNe Ibc. However, in this case the purity of output SN II sample is very high. In fact, by using a combined classifier consisting only of \dash{} and \ngsf{} we retrieve some of the classification completeness, classifying just under a third of SNe II successfully to produce a sample that is 96.4\% pure. Similarly, one can obtain a 77\% pure sample of SNe Ibc, although this can be improved to 92\% at the cost of only one-third of the completeness (44\% to just 17\%) if \dash{}--\snid{} is used instead.

Due to \dash{}'s presence in this combined classifier, the classification completenesses of SLSNe and non-SN transients are zero. Indeed this can also be seen in Fig. \ref{sing doub cm}, in both double classifier combinations including \dash{}, which cannot output SLSN classifications without retraining with a different template set that contains SLSN spectra. 

The poor classification completeness shown in Fig. \ref{sing doub cm}(a) and Table \ref{combined puritys} suggests that the use of combined classifiers alone is not particularly appropriate for live transient classification. However, it does indicate the potential for very pure SN Ia and SN II samples, although the latter sample has very low classification completeness. As a result, combined classifiers could still form an important part of a live classification plan. 

A combined classifier could be used as a first classification step to remove this high purity SN Ia sample prior to additional, later classification steps. Depending on the classifier used, this can also be done for the very pure (but low completeness) SN II sample produced. When spectroscopic redshifts are known, \dash{}--\ngsf{} is an obvious choice due to its high \(F_{0.5}\)--score. Without redshifts it should be noted that a \dash{}--\snid{} classifier returns the best \(F_{0.5}\)--score. The marginally reduced purity is compensated by the higher completeness. However, unlike the case of known redshifts where \dash{}--\ngsf{} is clearly the best performing classifier, when redshifts are not known all three double classifiers have similar $F_{0.5}$--scores. Both with spectroscopic redshifts and unknown redshifts, when using all three classifiers, the reduction in completeness is more significant than the negligible improvement in purity compared to classifying with \dash{}--\ngsf{} only. We investigate the potential for a second stage of classification in Section \ref{ppcs}.

\begin{table*}
    \centering
    \caption{The SN Ia completeness and purity for all possible combinations of two or three classifiers. Successful classification requires a SN Ia output from all involved classifiers. For the combined \ngsf{}--\snid{} classifier we also report the same results assuming the presence of photometric priors. The highest value in each column is highlighted in bold.}
    \begin{tabular}{c|c|c|c|c}
        Classifiers & Redshift & Ia completeness & Ia Purity & \(F_{0.5}\)-Score \\
        \hline
        \dash{} \& \ngsf{} & Known \(z\)& \textbf{0.689$\pm$0.005} & 0.9995$\pm$0.0003 & \textbf{0.757 $\pm$ 0.004} \\
        \ngsf{} \& \snid{} & . & 0.621$\pm$0.006 & 0.9994$\pm$0.0003 & 0.687 $\pm$ 0.005\\
        \dash{} \& \snid{} & . & 0.623$\pm$0.006& 0.9984$\pm$0.0004 & 0.674 $\pm$ 0.006\\
        ALL & . & 0.590$\pm$0.006 & \textbf{1.0$\pm$0.0} & 0.669 $\pm$ 0.005\\
        \hline
        \dash{} \& \ngsf{} & Unknown \(z\) & 0.367$\pm$ 0.004 & 0.997$\pm$0.001 & 0.566 $\pm$ 0.006 \\ 
        \ngsf{} \& \snid{} & . & 0.424$\pm$0.006 & 0.976$\pm$0.004 & 0.566 $\pm$ 0.006\\ 
        \dash{} \& \snid{} & . & \textbf{0.456$\pm$0.006} & 0.991$\pm$0.001 & \textbf{0.589 $\pm$ 0.006}\\
        ALL & . & 0.324$\pm$0.005 & \textbf{0.998$\pm$0.001} & 0.510 $\pm$ 0.006\\
        \hline
        \ngsf{} \& \snid{} & Photo-\(z\) & 0.427 $\pm$ 0.007 & 0.990 $\pm$ 0.001 & 0.553 $\pm$ 0.004\\
    \end{tabular}
    \label{combined puritys}
\end{table*}

We conclude that that the best performing classifier is \dash{}--\ngsf{}. When redshifts are known, the SN Ia and SN II completenesses is 10 percentage points higher or more than using all three classifiers. This amounts to the addition of hundreds of transients into the final sample at the cost of doubling an already negligible non-SN Ia contamination. In the case where redshifts are not known this logic holds true, but with a combination of \dash{} and \snid{}. As shown in Table \ref{5class_comp}, \ngsf{} is particularly affected by a lack of redshift information. However, without redshift priors, all three double classifiers perform similarly with regard to $F_{0.5}$--scores.

\subsection{Potential Photometric Cuts} \label{ppcs}

Individually, we see mixed results from the classifiers. Depending on the classifier and redshift information used, completeness can change by up to 50\% and SN Ia purities by as much as 15\%. From a cosmology perspective we obtain both high-purity and reasonably high completeness in SN Ia classification from \dash{} and \ngsf{}, but only when redshift information is known, and it is yet unclear to what extent prior redshift information will be available for TiDES transients.

From a live classification perspective, there appears to be no single classifier from which we can expect a reasonable classification completeness across the SN Ibc, II, SL and non-SN classes. More importantly, the result of these low completenesses is that misclassified transients must be contributing to lowering the purity of some other class.

To this point we have attempted classification on every transient that has received any exposure time in the survey simulation. We will now investigate two obvious sources of `other' classification to see if applying cuts to the sample prior to classification will improve results. In Section \ref{TFF cuts} we investigate making cuts on the fraction of fibre flux deriving from the transient (as opposed to its host galaxy) and in Section \ref{Smag cuts} we investigate cuts based on the brightness of the transient. Both of these quantities should be reasonably obtainable from the same LSST photometry that TiDES will use to flag potential transient targets. 

In both cases, photometric cuts are performed based on the LSST \(r\)--band magnitude at the time of simulated 4MOST observation. The transients in the simulation are binned in phase every five days and so there may be a discrepancy between of a few days between the simulated observation and the date of the reported magnitude. In reality, transients added to the 4MOST observing queue, for which we know the triggering magnitude from LSST, will only remain in the 4MOST observing queue for four days \citep{2025arXiv250116311F} before needing refreshed with fresh photometry. So a discrepancy of several days between last known magnitude and 4MOST observation is realistic. We expect transient alert packets from LSST to be sufficient to perform the following photometric cuts.

\subsubsection{Apparent Transient Magnitude} \label{Smag cuts}

The most obvious sample cut that can be introduced from photometric information is a cut on transient magnitude. In this section we investigate the potential for applying a cut to our transient sample based on the \(r\)-band magnitude of the transient.

\begin{figure}
    \centering
\includegraphics[width=\linewidth]{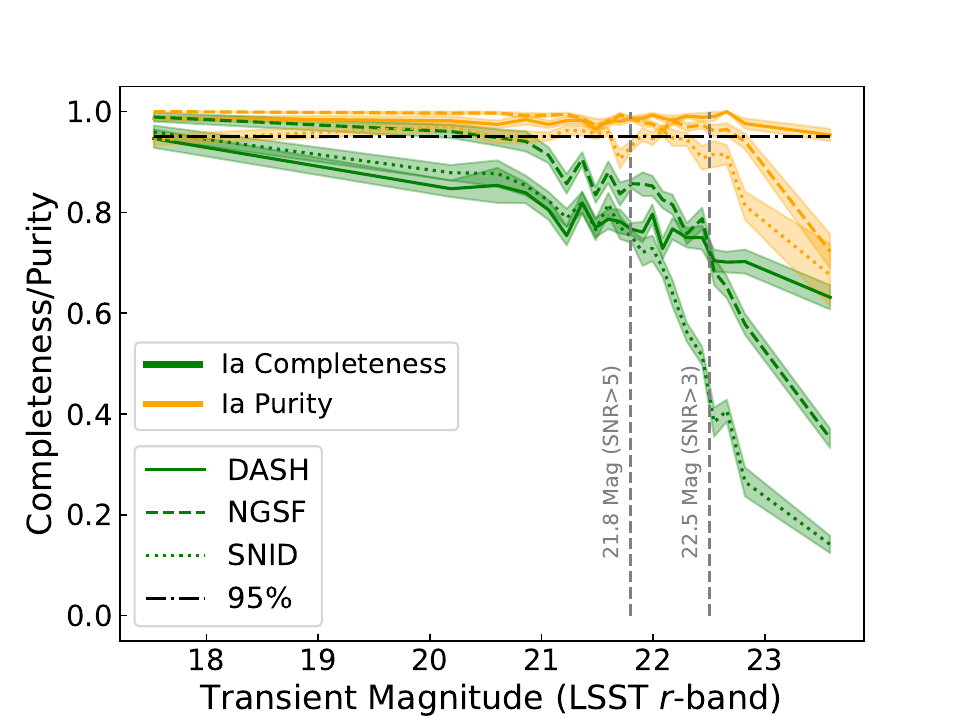}
    \caption{The SN Ia purity (orange) and completeness (green) as report by \dash{}, \ngsf{} and \snid{} as a function of the true transient magnitude for the SNe Ia in all of our subsamples. The SNe Ia are in non-linear magnitude bins of $\sim$30 transients, with each plotted point at its bin's centre. The shaded areas indicate the standard error on the mean of completeness and purity in each bin. 95\% purity is marked by a black dashed line. Two potential transient magnitude cuts are marked by grey dashed lines at 21.8 and 22.5 mag. We find that these limits roughly correspond to completeness dropping below 80\% and purity falling below 95\%, respectively.}
    \label{smag success}
\end{figure}

Fig. \ref{smag success} presents the completeness and purity of SN Ia classification for all three classifiers as a function of transient r-band magnitude. It also proposes two potential values for a transient magnitude cut to our sample. These values, 21.8 and 22.5 mag, are derived in \citet{2025arXiv250116311F} as the magnitudes that correspond to transient spectral SNRs of 5 and 3, respectively, where spectral SNR is calculated as the average in 15 \AA \ bins between 3,500 and 8,000 \AA.  Indeed \citet{2025arXiv250116311F} reports the SNR = 5 threshold as the conservative minimum to meet TiDES's spectral success criteria, with the SNR = 3 limit a more optimistic estimate based on the work of \citet{2009A&A...507...85B}. Here, we find that these SNR cuts of 5 and 3 correspond roughly to the SN Ia completeness falling below 80\% and the purity falling 95\%, respectively.

\begin{table*}
    \centering
    \caption{Ia classification results and 5-class weighted \(F_{0.5}\)--score for \ngsf{}. We report the results with \(r\)-band magnitude cuts of 21.8 and 22.5 mag, as well as with no cuts. Completeness and \(F_{0.5}\)--score are calculated with the sample size \textit{after} the cut is applied, but we note that mean Ia sample is reduced in size to 55\% and 83\% by magnitude cuts at 21.8 and 22.5 mag, respectively. The highest values in each column are highlighted in bold.}
    \begin{tabular}{c|c|c|c|c}
        Redshift Prior & \(r\)-band Cut & Ia Completeness & Ia Purity & \(F_{0.5}\)--Score \\
        \hline
        Known \textit{z} & 21.8 & \textbf{0.882 $\pm$ 0.005} & \textbf{0.987 $\pm$ 0.002} & \textbf{0.876 $\pm$ 0.003}  \\
         . & 22.5 & 0.837 $\pm$ 0.005 & 0.981 $\pm$ 0.002 & 0.842 $\pm$ 0.003 \\
        . & None & 0.798 $\pm$ 0.005 & 0.971 $\pm$ 0.002 & 0.814 $\pm$ 0.004\\
        \hline
        Unknown \textit{z} & 21.8 & \textbf{0.606 $\pm$ 0.006} & \textbf{0.936 $\pm$ 0.005} & \textbf{0.668 $\pm$ 0.006}\\
         . & 22.5 & 0.585 $\pm$ 0.006 & 0.933 $\pm$ 0.005 & 0.655 $\pm$ 0.005 \\
        . & None & 0.560 $\pm$ 0.006 & 0.917 $\pm$ 0.005 & 0.627 $\pm$ 0.005\\
         \hline
    \end{tabular}
    \label{binary smagcut res}
\end{table*}

As \ngsf{} produced the best individual \(F_{0.5}\), in Table \ref{binary smagcut res} we present classification results from \ngsf{}, but now with the effects of cutting transients fainter than 21.8 and 22.5 mag. This does remove nearly half of the transients from the final sample for the stricter 21.8 mag cut. However, we generally see significant improvements across SN Ia completeness, SN Ia purity and \(F_{0.5}\)--score as stricter magnitude cuts are employed.

\dash{} and \snid{}, while not shown, also follow this trend. \ngsf{} outperforms \snid{} across all metrics both with and without redshift priors. However, without redshifts \dash{} does produce $F_{0.5}$--score about 0.01 larger than \ngsf{}, mainly the result of \dash{} maintaining a high Sn Ia purity which is very heavily weighted in the $F_{0.5}$--score. However, \ngsf{}, with spectroscopic redshifts, produces $F_{0.5}$--scores around 0.1 larger than \dash{} or \snid{}.

Cutting on \(r\)-band magnitude results in a significant reduction in sample size, so this would not be appropriate by itself for automatic classification. However, it could serve as a useful step in a pipeline for broad classification.

In Section \ref{multiple} we found that, while combined classifiers are very good at creating high purity, low completeness SN Ia samples, they are poor classifiers of non-SN Ia classes. This makes them ineffective for TiDES live transient classifications. We also found in Sections \ref{ind results} and \ref{5 class}, that the individual classifiers produce mediocre completeness and purity in most transient classes when operating on every transient observed in the 4MOST survey simulation. However, for TiDES transients brighter than \(r\)=21.8 mag, \ngsf{} appears to be a good choice for automated live classification.

However, this comes with several caveats. First, there will be significant performance loss when redshift information cannot be provided.  Second, this only applies with relatively broad transient classes. For example, \ngsf{} often classifies Ib-norm inputs as SN Ic subclasses. Just under 50\% of Ibc classification are SNe Ib classified as SNe Ic and vice versa. Finally, and perhaps most importantly, while the SNe Ia purity is high, the purity of the other classification bins can be far lower. For example the SN II purity is 77\%, and the Ibc purity is just 70\% (see Table \ref{ngsf cuts}).

From the point of view of the potential cosmology sample of SNe Ia obtained in Section \ref{multiple}, cutting transients from our sample based on their apparent magnitudes has less impact on the purity than the completeness. All three classifiers see between 0-4\% improvement. Compared to the needs of live classification, it is less clear if this small improvement in purity compensates for the significant fraction of the sample discarded before classification. In fact, the \dash{}--\ngsf{} combined classification produces a higher SN Ia purity and classifies a greater number of SNe Ia in total (since the completeness of the 21.8 mag cut \ngsf{} classification is around 50\% when cut transients are accounted for).

\subsubsection{Transient Flux Fraction} \label{TFF cuts}

After transient magnitude, the second obvious source of classification error in our sample comes from high levels of host galaxy flux in our spectra. In this section we discuss the effectiveness of \dash{}, \ngsf{} and \snid{} as a function of transient flux fraction (contrast), where the transient flux fraction is the fraction of the flux in a 4MOST fibre that originates from the transient. We report the potential to improve classification results by introducing a sample cut in transient flux fraction-redshift space. We investigate using our 5-class classification schema as in previous sections.

\begin{figure}
\centering
\includegraphics[width=\linewidth]{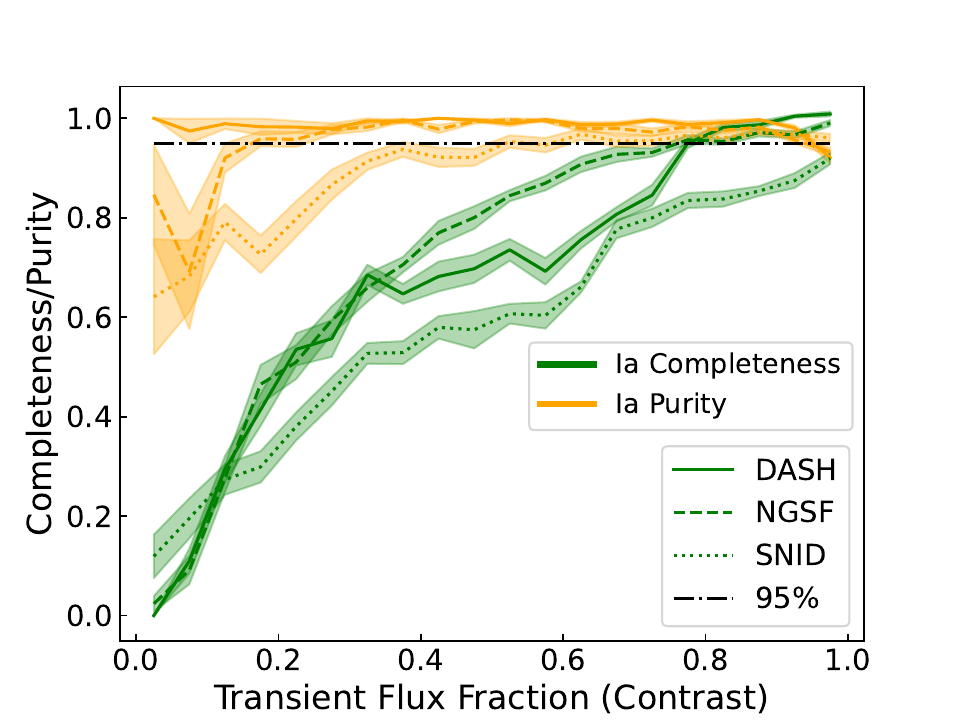}
\caption{The SN Ia completeness (green) and purity (orange) as a function of the fraction of the total flux in the spectrum that originates from the transient. The SNe Ia in each of our subsamples are in 20 linear bins between transient fibre flux fractions (contrast) of 0 and 1. Redshift is known in all cases. Uncertainty in indicated by the shaded regions. Shaded regions are defined by the standard error on the mean in each bin between our random subsamples. All three classifiers produce similar trends in SN Ia completeness and purity. In every case the classification completeness and purity improve as the transient flux fraction increases.}
\label{contrast_plots}
\end{figure}

Generally, the trends in classification rates against the transient flux fraction are as one would expect. As the transient flux fraction increases (the spectrum's host contamination is reduced) we see improvements in the SN Ia completeness and purity. The shape of these plots is very similar to those produced by transient magnitude binning in Fig. \ref{smag success}. The purity tends to approach 95\(\%\) at transient flux fractions of 40 - 50\(\%\) if it is not already above that in the most contaminated bin. Fig. \ref{contrast_plots} indicates that all three classifiers have similar slopes in their purity with different initial values. Although not shown in the figure, the same trend was found without redshift priors, albeit with slightly smaller values for \dash{} and much smaller values for \ngsf{} and \snid{}.

\begin{figure*}
    \centering
\includegraphics[width=\linewidth]{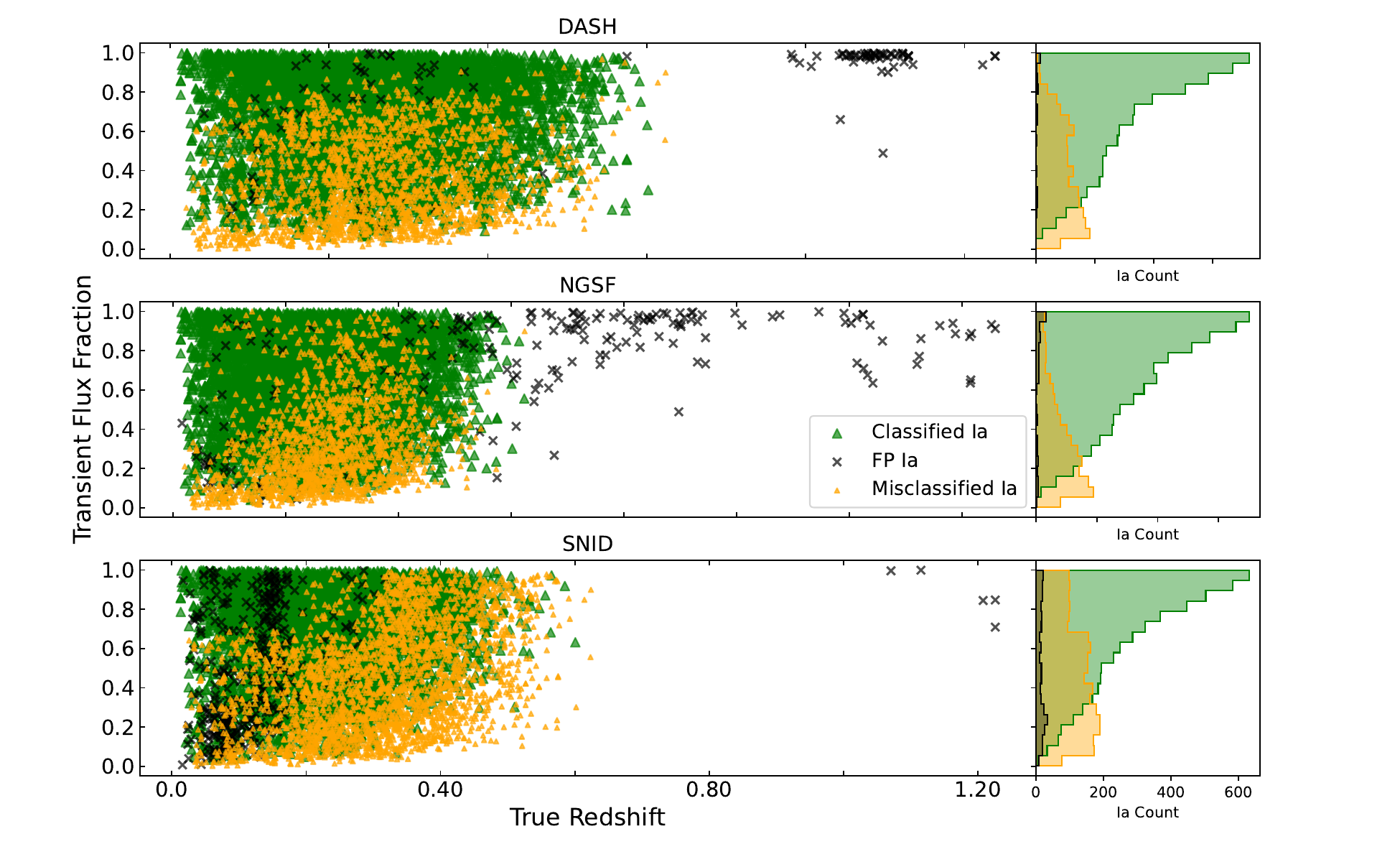}
    \caption{The classification results in the binary schema with known redshifts for all three classifiers in transient flux fraction-redshift space. Green and orange points indicate good SN Ia classifications and failed SN Ia classifications, respectively. The black crosses indicate SN Ia false positives (that is, a non-SN Ia classified as a SN Ia.) The histograms show the corresponding counts with the same colour scheme. There are regions of the parameter space for each classifier where false positive SN Ia classifications cluster, often at high redshifts (\(z > 0.6\)). We also see similar distributions for successful and unsuccessful SN Ia classifications.}
    \label{param space}
\end{figure*}

We look at our results in flux fraction-redshift space in Fig. \ref{param space}. At high redshift only transients that have bright absolute magnitudes, especially transients in the SLSN class, will be observed. So transient flux fraction is likely to be high as we are biased to intrinsically brighter transients while host brightness remains constant.  However, we also expect the spectral features of our transients to be shifted outside of 4MOST's wavelength range, making them harder to classify. Indeed the \(rlap\) classification quality parameter employed by \dash{} and \snid{} depends directly on the wavelength overlap between the input spectrum and matching template. We hope to find regions of this parameter space without contaminants or fewer misclassifications, where we could assign positive results a greater degree of certainty.

A few obvious points of interest are the trend to greater transient flux fractions with increasing redshift and the incidence of unsuccessful classifications of SNe Ia (orange histograms) beginning to drop off as the transient flux fraction surpasses around 40\%. The SN Ia count histograms are fairly uniform for the three classifiers in the relative distributions of the successful and unsuccessful SN Ia classifications, but we see variation in the width of the successful classification histogram. In particular, there are obvious differences in the number of misclassified SNe Ia between the classifiers.  

Also concerning are the clusters of SLSNe at high redshift (\(z > 0.6\)) that are classified as SNe Ia in all three classifiers, although most prevalently in \dash{} and \ngsf{}. These SLSNe are being fit overwhelmingly as SNe Ia-91bg. This does lead to a potential mechanism for increasing purity. As can be seen in Fig. \ref{param space}, the successful SN Ia classifications (and indeed instances of SNe Ia in general) drop off quite sharply after \(z = 0.60\). Each classifier has contaminants beyond this redshift that could be dismissed out of hand if accurate spectroscopic redshifts for host galaxies are known, or if photometric redshifts indicate it is likely that \(z > 0.60\).

For now, with the precise extent to which TiDES will have host redshift information, we do not implement such a cut. However, we make note of it and strongly encourage such a cut's usage in the cases where redshifts are known.

An obvious location for a cut on the transient flux fraction is the point at which the good SN Ia classifications begin to dominate over misclassifications. This occurs at a transient flux fraction of roughly 0.2 for \dash{}, 0.2 for \ngsf{} and 0.3 for \snid{}, we generalise this to a cut at a flux fraction of 0.3.

A second tempting cut is on very large transient flux fractions, greater than 0.9. In \dash{} and \ngsf{} there are clusters of very bright, high flux fraction, SLSNe being falsely classified as SN Ia. However, we choose not to pursue this cut, simply because removing SNe in these bins would also remove the regions with the highest density of correct classifications.

\begin{table*}
    \centering
    \caption{The completeness and purity of each of our classes in the 5-class scheme under photometric cuts. The magnitude cut requires SNe \(r\)-band magnitude < 21.8 and reduces the sample size to 61.7\%. The flux fraction cut requires that transient flux fraction > 0.3 and reduces the sample size to 80.3\%. Using both reduces the sample size to 52.2\%. Completeness and \(F_1\)--score are the based on the transients in the classified sample, so objects removed by the photometric cuts do not contribute. Only \ngsf{} is shown, having been identified as the most promising candidate for live classification.}
    \begin{tabular}{c|c|c|c|c}
    Metric & No Cut & Mag. Cut & Flux Frac. Cut & Both \\
    \hline
    Ia Comp. & 0.798 $\pm$ 0.005 & 0.882 $\pm$ 0.005 & 0.888 $\pm$ 0.004 & \textbf{0.952 $\pm$ 0.003}\\
    Ia Purity & 0.971 $\pm$ 0.002 & 0.987 $\pm$ 0.002 & 0.973 $\pm$ 0.002 & 0.\textbf{988 $\pm$ 0.002}\\
    \hline
    Ibc Comp. & 0.52 $\pm$ 0.02 & 0.65 $\pm$ 0.02 & 0.64 $\pm$ 0.02 & \textbf{0.75 $\pm$ 0.02}\\
    Ibc Purity & 0.61 $\pm$ 0.02 & 0.70 $\pm$ 0.02 & 0.72 $\pm$ 0.02 & \textbf(0.84 $\pm$ 0.02)\\
    \hline
    II Comp. & 0.753 $\pm$ 0.006 & 0.836 $\pm$ 0.009 & 0.78 $\pm$ 0.01 & \textbf{0.88 $\pm$ 0.01}\\
    II Purity & 0.748 $\pm$ 0.009 & 0.767 $\pm$ 0.007 & 0.847 $\pm$ 0.007 & \textbf{0.860 $\pm$ 0.007}\\
    \hline
    SL Comp. & 0.85 $\pm$ 0.01 & \textbf{0.913 $\pm$ 0.007}& 0.845 $\pm$ 0.006 & \textbf{0.913 $\pm$ 0.007}\\
    SL Purity & 0.51 $\pm$ 0.01 & 0.75 $\pm$ 0.01 & 0.62 $\pm$ 0.01 & {0.84 $\pm$ 0.02}\\
    \hline
    Non-SN Comp. & 0.05 $\pm$ 0.02 & \textbf{0.07 $\pm$ 0.02} & 0.04 $\pm$ 0.02 & 0.04 $\pm$ 0.02\\
    Non-SN Purity & 0.04 $\pm$ 0.01 & 0.05 $\pm$ 0.02 & \textbf{0.15 $\pm$ 0.08} & 0.13 $\pm$ 0.08\\
    \hline
    \(F_{0.5}\)--score & 0.814 $\pm$ 0.004 & 0.876 $\pm$ 0.003 & 0.866 $\pm$ 0.003 & \textbf{0.920 $\pm$ 0.002} \\

    \end{tabular}
    \label{ngsf cuts}
\end{table*}

In Table \ref{ngsf cuts} we present the results of our 5-class classification schema for \ngsf{} as we employ a variety of different photometric cuts to the input sample. We see that using only a cut for transient flux fractions greater than 0.3 returns similar classification results across most transient classes to the 21.8 transient magnitude cut employed in Section \ref{Smag cuts}. The Ia classification performance is nearly identical, with the other classes best performances spread fairly evenly. Using both cuts results in even better performance, indeed it produces the largest \(F_{0.5}\)--score, followed by the magnitude cut and then the flux fraction cut. However, these performance benefits must be weighed against the large fractions of the sample removed from consideration and thus not reflected in the \(F_{0.5}\)--score.

We conclude cautiously that the best photometric cut for live classification is likely to be transient transient magnitude \(r\)>21.8, the middle ground between improved performance and reduction in sample size. Although arguments can be made for the flux fraction cut or both. In all three cases the non-SN transient completeness and purities are very poor. This is the result of low numbers (or a complete absence) of templates in the template banks/training samples and, additionally, the fact that non-SN input spectra are just smooth-blue continua (see Appendix \ref{example spec}).

\subsection{An Example Classification Plan} \label{an example}

\begin{figure*}
    \centering
\includegraphics[width=0.6\linewidth]{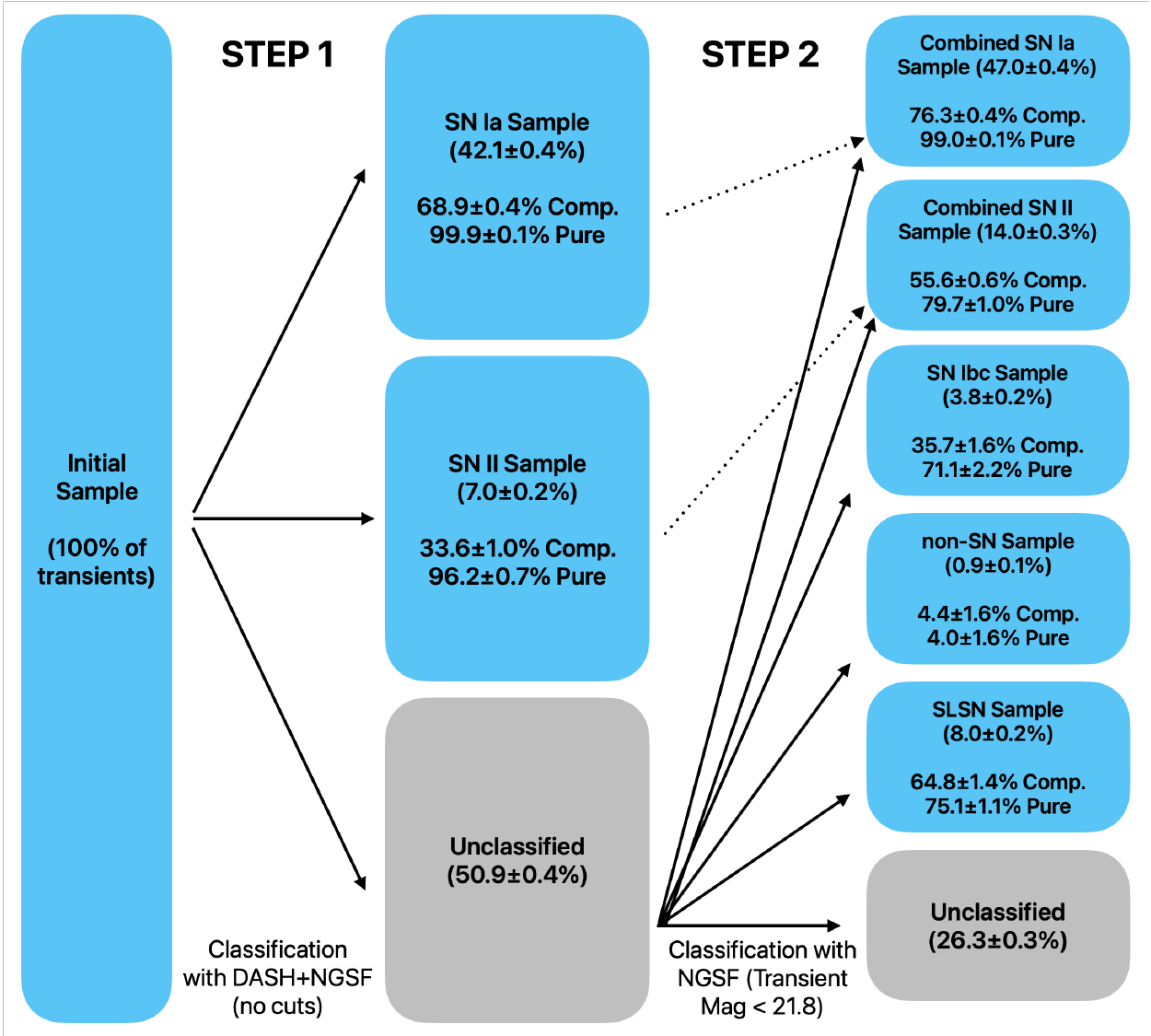}
    \caption{An example of a classification pipeline that could be employed by TiDES for the purpose of live classification of transients. The output samples of from each step in the classification pipeline are provided with their completeness and purities labeled. The samples of SNe Ia and SNe II provided after the second classification step represent the combination of the transients from the first classification and those from the second. Percentages of the total sample size are listed in brackets for each classes final sample. Classifications are performed with redshift information.}
    \label{pipeline}
\end{figure*}

In this section we propose just one possible scheme that could be employed by TiDES for live classification of transients. The pipeline is illustrated in Fig. \ref{pipeline} and assumes redshift information is provided for all classifications. The pipeline consists of two separate classifications of the sample of transients. First, the full sample is classified by the combined \dash{}-\ngsf{} classifier recommended in Section \ref{multiple}. This produces very pure samples of SNe Ia and SNe II although, particularly for the latter, the completeness is low. The SNe Ia sample produced by this first classification step has 99.9\% purity and should be appropriate for use in cosmology.

From the sample of spectra not classified by the combined classifier, we now take only those with a transient magnitude brighter than 21.8 mag as discussed in Section \ref{Smag cuts}. These bright objects are then reclassified with just \ngsf{}. This produces reasonably pure and complete samples of SNe Ibc and SLSNe. It also classifies a few additional SNe Ia and SNe II which can be combined with the existing samples to increase their completeness at the cost of their purities. The only class with poor results is the non-SN transients. Here we only classify 4\% correctly and over 95\% of the resulting sample is contamination from other classes. This is an issue with \ngsf{}'s template bank and the absence of such spectra from \dash{}'s training set. When considered in full, the classification pipeline leaves just over a quarter of transients unclassified.

This is a reasonably successful classification. It outperforms any individual spectroscopic classifier that we have tested in this work in regards to purity. This classification scheme obtains a very pure SNe Ia sample for cosmology in addition to producing classification completeness and purities in non-SN Ia classes that are suitable for live transient classification. See Fig. \ref{pipeline} for the completeness and purity of each class after each step of the classification pipeline.

We note that this classification pipeline has a higher \(F_{0.5}\)--score than \ngsf{}. However, the choice of $\beta$ in Equation \ref{fbeta} allows for greater importance to be placed on completeness rather than purity. The \(F\)--scores for several values of $\beta$ across several classification schemes are presented in Table \ref{fscores}. We can see that while \ngsf{} individually performs best in \(F_1\) and \(F_2\)--score, when the score is weighted to favour completeness (\(\beta>1\), the various versions of the classification pipeline presented in this section have the highest score when $\beta=0.5$ and purity is weighted more heavily. In fact, at even lower values of \(\beta \leq 0.1\), the combined \dash{}--\ngsf{} classifier would have the best score. As a result, it is hard to objectively state the superior classifier, it will depend on the objectives of a particular study.

Fortunately, there is significant room for fine-tuning to specific science cases. For example, replacing the cut on transient magnitude to the cut on transient flux fraction as discussed in Section \ref{TFF cuts}, the pipeline will produce samples with higher completeness at the cost of purity. Additionally the percentage of unclassified objects drops to just 18\%. In this case the SLSN purity drops to around 65\%, but this is compensated by an completeness of over 80\%.

\begin{table}
    \caption{The \(F_{0.5}\), \(F_{1}\) and \(F_{2}\) scores of several classifiers mentioned throughout this paper. Each choice of $\beta$ indicates a different priority in the classifier. Smaller $\beta$ values increasingly weight the \(F\)--score towards good purity results, while increasingly large values instead weight in favour of completeness. $\beta$ values of 0.5 and 2 and used by convention. The largest value(s) in each column are in bold.}
    
    \centering
    \begin{tabular}{c|c|c|c}
        Classifier & \(F_{0.5}\) & \(F_{1}\) & \(F_{2}\) \\
        \hline
        \makecell{Pipeline: \\ Mag. Cut} & \textbf{0.830 $\pm$ 0.002} & 0.757 $\pm$ 0.002 & 0.698 $\pm$ 0.003 \\
        \hline
        \makecell{Pipeline: \\ Flux Frac. Cut} & \textbf{0.831 $\pm$ 0.003} & 0.773 $\pm$ 0.003 & 0.726 $\pm$ 0.003 \\
        \hline
        \dash{}--\ngsf{} Only & 0.757 $\pm$ 0.004 & 0.645 $\pm$ 0.005 & 0.566 $\pm$ 0.004 \\
        \hline
        \ngsf{} Only & 0.814 $\pm$ 0.004 & \textbf{0.786 $\pm$ 0.004} & \textbf{0.765 $\pm$ 0.004} \\
    \end{tabular}
    \label{fscores}
\end{table}

Additional cuts from photometric information can be added to either stage of the pipeline to increase purity at the cost of completeness. Different cuts than those discussed here can be used, which will affect each class differently, allowing for parties interested in specific SNe classes to be specific in their classification.

The final advantage of such a classification model is that it is versatile and easily communicated to the community. By providing only the class from the 5-class output probabilities from each classifier, the \(r\)-band magnitude of the transient and host near time of observation and the redshift of the system, it would be possible for members of the community to adjust the transient sample selected to suit their specific science requirements by varying classifiers or probability thresholds.

\subsubsection{Comparison to Photometric Classification Results}

In this subsection, we compare three recent photometric classification papers surrounding a recent photometric classifier and its use with the Dark Energy Survey \citep{2022MNRAS.514.5159M}.

\citet{2020MNRAS.491.4277M} presents the photometric transient classifier \snnova \ classifying simulated light curves with spectroscopic redshift information and incomplete light curve information. Additionally, \citet{2022MNRAS.514.5159M} and \citet{2024MNRAS.533.2073M} present \snnova \ classification results on real light curves with and without host redshifts, respectively.

Specifically, \citet{2024MNRAS.533.2073M} presents the binary classification of DES 5-year data release SNe without any redshift information provided as a prior. When the light curves of transients being fit without redshifts are trimmed to only include photometry up to peak brightness, \snnova \ produces a binary accuracy, a Ia completeness and a Ia purity of 90.46 \%, 92.49 \% and 91.93 \%, respectively. By comparison, if operated as a binary classifier without redshift, our classification plan from Section \ref{an example} produces a binary accuracy, a Ia completeness and a Ia purity of 85.6$\pm$0.4 \%, 44.5$\pm$0.6\% and 94.4$\pm$0.3\%. Additionally, we can consider only the high-confidence SN Ia sample produced by the combined \ngsf{}-\dash{} classifier to improve the SN Ia purity to 99.5$\pm$0.1 \% at the cost of reducing completeness to just 36.4$\pm$0.6\%.

As seen in Table \ref{5class_comp}, \ngsf{} has significant performance loss when redshift information is not provided. As such, the binary accuracy, SN Ia completeness and purity can be improved to 91.4$\pm$0.4\%, 55.6$\pm$0.6\% and 95.3$\pm$0.3\% by replacing the \dash{}--\ngsf{} classification step with an equivalent \dash{}--\snid{} classification. However, this does come at the cost of worse performance in the 5-class mode of operation.

\citet{2022MNRAS.514.5159M} also applies \snnova \ to the photometric sample produced by the DES 5-year data release. This produces a cosmologically useful sample of 1,484 SNe Ia with spectroscopic redshifts. The predicted completeness and purity of the sample are 98.51\% and 97.73\%, respectively. Again, we consider both the high-confidence SN Ia sample and the larger, less confident, SN Ia sample produced by our classification pipeline. Now with redshift priors, the less confident sample has an completeness of 76.3$\pm$0.4\% and purity of 99.0$\pm$0.1\%. We can sacrifice some completeness to improve purity and use the high confidence SN Ia sample produced by the combined \dash{}-\ngsf{} classifier. This increases purity to >99.9\% with completeness just under 70\%. Our classification plan produces a SNe Ia sample with a percentage contamination that is more than a factor of ten lower, at the cost of lower completeness and accuracy, than \snnova. This is true whether redshift information is available or not.

While most photometric classifiers function purely in a binary (SN Ia v non-SN Ia) schema and with complete light curves, in \cite{2020MNRAS.491.4277M}, \snnova \ reports results using ternary and seven-way classification schema, similar to our 5-class schema.

\snnova \ reports an accuracy of 77.8\% for its ternary schema (SNe Ia, Ibc and II) and 64.2\% for the seven-way classification schema (SNe Ia, IIP, IIn, IIL1, IIL2, Ib, and Ic). In each case these are the accuracies expected from light curves consisting, on average, of 2.4 distinct nights of multi-colour observations up to 2 days before peak brightness. These percentages improve to 81.5\% and 69.8\% for an average of 3.1 distinct nights of multi-colour observations up to 2 days after peak brightness. All classifications also make use of spectroscopic redshifts.

For comparison our example pipeline, in the 5-class schema (SNe Ia, Ibc, II, SL and non-SNe), produces a comparable classification accuracy of 90.1$\pm$0.2\%. Additionally, if we consider only SNe Ia, Ibc and II to mimic the ternary schema, we obtain an accuracy of 93.2$\pm$0.3\%. In both cases we don't consider unclassified spectra in our calculation of the accuracy. In the ternary scheme, non-SN transient and SLSN outputs are considered unclassified.

From \citet{2025arXiv250116311F} the requirements to flag a transient for spectroscopic follow-up are 3 \(griz\) detections in two distinct nights, with the added requirement that at least one of these detections be brighter than 22.5 mag. We also assume spectroscopic redshifts are available. Our use of spectroscopy produces a roughly 15\% improvement on the accuracies from photometry with similarly incomplete light curves.

\section{Conclusions} \label{conc}

In this paper we set out to determine whether the classification of transients discovered by 4MOST-TiDES can be automated using one or more spectroscopic transient classifiers. We want to know which classifier(s) are the best from a live-classification and cosmological point of view. To do this, we simulated realistic blended spectra using pre-existing simulations and the 4MOST ETC and classified them using \dash{}, \ngsf{} and \snid{}. We place a focus on classification purity due to the large sample sizes produced by TiDES, and employ the \(F_{0.5}\)--score as our purity-weighted FoM.

The classification performances of \dash{}, \ngsf{} and \snid{} are weaker than those reported in their original papers. This is the result of different quality data and fainter SNe, alongside significant host contamination. We find that, individually, \ngsf{} produces the best \(F_{0.5}\)--score for known redshift classifications, although its performance loss is across all transient classes large if redshift information cannot be provided. None of the individual classifiers were robust enough to recommend their use for automated classification.

We find that the purities in SNe Ia can be greatly improved by using several classifiers at once and requiring an agreement between them on each classification. This is costly for transient completeness, but with the benefit of having vastly reduced contamination in the output sample. We get good results from a combination of \dash{} and \ngsf{}, with SNe Ia completeness of \(69.4 \pm 0.5\)\% and purity of \(99.94 \pm 0.03\). Purity can be marginally improved by including \snid{} in the combined classifier, but at the cost of a much reduced completeness.

This allows for the automation of SNe Ia classification and the production of good cosmology samples. However, it alone does not lead to a solution for general automated classification for TiDES. The combined \dash{}-\ngsf{} classifier struggles to classify SNe Ibc and SNe II with a high completeness, although what it does classify is quite pure. It is incapable of classifying SLSNe and non-SN transients, as \dash{}, by default, has not been trained to classify them.

We investigated a variety of photometric cuts that could be applied to our data to improve the resulting transient classifications for individual classifiers. We found that only classifying transients with \(r\)-band magnitudes brighter than 21.8 could significantly improve classification purity across all transient classes, but at the cost of classification completeness. Similar results can be obtained by only classifying objects for which SNe flux comprises more than 30\% of the flux within the observing 4MOST fibre.

We present an example classification plan in Section \ref{an example}. We emphasize that such a classification pipeline is easily fine-tuned to specific science cases and conclude it is viable for live automated classification and these modifications require only the classifier outputs and some photometric information to be performed. The specific classification pipeline present in this paper outperforms the \(F_{0.5}\)--scores of all combinations of one, two or three classifiers. In Table \ref{fscores} we indicate how one might choose a different classifier than our pipeline depending on whether the purity of the sample or the completeness is considered most important for particular research goals.

We have demonstrated the capacity of an example classification pipeline to produce a very high purity SN Ia sample at the cost of completeness, and a sample with far higher completeness with lower purity. A future step in this work will be to optimize the classification scheme via end-to-end cosmological simulations, in order to show which sample best constrains the cosmology and which combination of classifiers and photometric cuts minimize the uncertainty on derived cosmological parameters.

Importantly, it is currently unclear to what extent 4MOST-TiDES will be able to obtain redshift information from host galaxies to be used in transient classification. The change in completeness and purities is significant between known and unknown redshifts and represents perhaps the largest uncertainty in the results of this paper. Work is currently underway investigating how consistently a redshift can be derived from features in blended host--transient spectra. Even in the case that live spectroscopic redshifts cannot be obtained from hosts, we are optimistic that it will be possible to obtain some host-redshifts from legacy surveys such as DESI \citep{2019AJ....157..168D} and SDSS \citep{2000AJ....120.1579Y, 2023ApJS..267...44A}. Host photo-\(z\)s also present a promising middle-ground between spectroscopic and unknown redshifts.

Finally, it is likely to be possible to bolster the spectroscopically confirmed transient samples with photometrically classified transients once full light-curve data is produced by LSST.

\section*{Acknowledgements}

AM gratefully acknowledges support from an STFC PhD studentship and the Faculty of Science and Technology at Lancaster University. IH gratefully acknowledges support from the Leverhulme Trust [International Fellowship IF-2023-027] and the Science and Technologies Facilities Council [grants ST/V000713/1 and ST/Y001230/1]. Y.-L.K. has received funding from the Science and Technology Facilities Council [grant number ST/V000713/1] and was supported by the Lee Wonchul Fellowship, funded through the BK21 Fostering Outstanding Universities for Research (FOUR) Program (grant No. 4120200513819) and the National Research Foundation of Korea to the Center for Galaxy Evolution Research (RS-2022-NR070872, RS-2022-NR070525). AM is supported by the ARC Discovery Early Career Researcher Award (DECRA) project number DE230100055. ET was supported by the Estonian Ministry of Education and Research (grant TK202), Estonian Research Council grant (PRG1006) and the European Union's Horizon Europe research and innovation programme (EXCOSM, grant No. 101159513). KM is funded by Horizon Europe ERC grant no. 101125877. PW acknowledges support from the Science and Technology Facilities Council (STFC) grants ST/R000506/1 and ST/Z510269/1. R.D. gratefully acknowledges support by the ANID BASAL project FB210003. MN is supported by the European Research Council (ERC) under the European Union’s Horizon 2020 research and innovation programme (grant agreement No.$\sim$948381) and by UK Space Agency Grant No.$\sim$ST/Y000692/1. The authors thank the anonymous reviewer for comprehensive comments that led to a much improved paper. The authors thank the anonymous referee for their useful comments.

\section*{Data Availability}
The set of \snid{} templates used throughout can be made available on request. Additionally, the full set of blended spectra used throughout are to be made available through a public repository on Lancaster University's PURE research information system.

\newcommand{\newblock}{}
\bibliographystyle{mnras}
\bibliography{class_comp.bib}

\nocite{*}

\appendix

\section{SNe Ia Fits and contaminant Origins}

\begin{figure*}
\begin{subfigure}{0.48\textwidth}
\includegraphics[width=\linewidth]{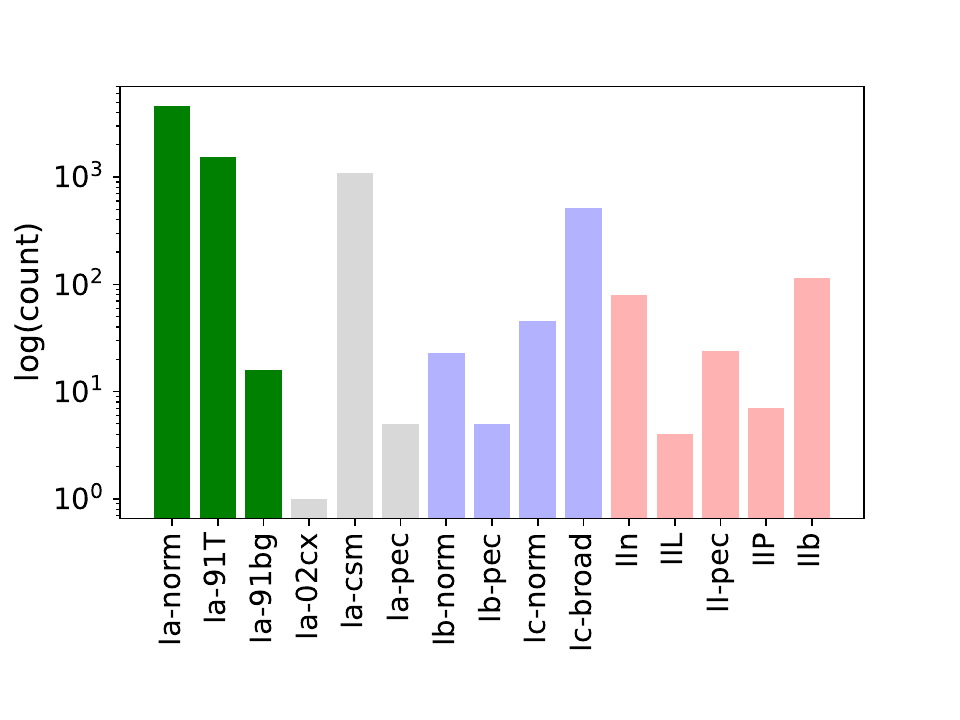}
\caption{\dash{} known \(z\)} \label{Ia:Dash known z}
\end{subfigure}\hspace*{\fill}
\begin{subfigure}{0.48\textwidth}
\includegraphics[width=\linewidth]{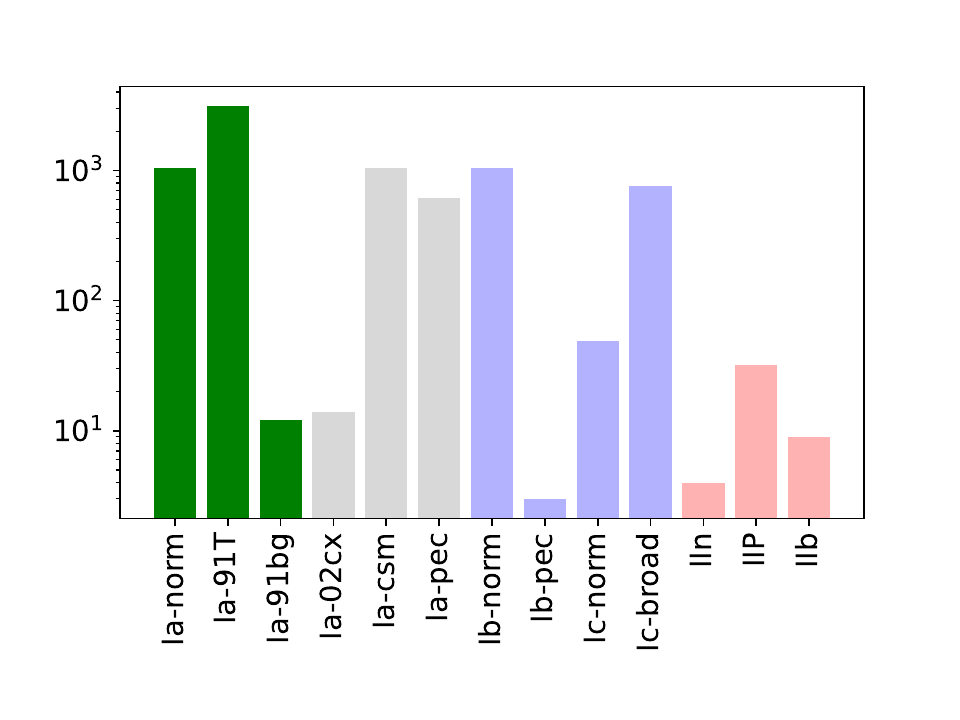}
\caption{\dash{} unknown \(z\)} \label{Ia:DASH unknown z}
\end{subfigure}

\begin{subfigure}{0.48\textwidth}
\includegraphics[width=\linewidth]{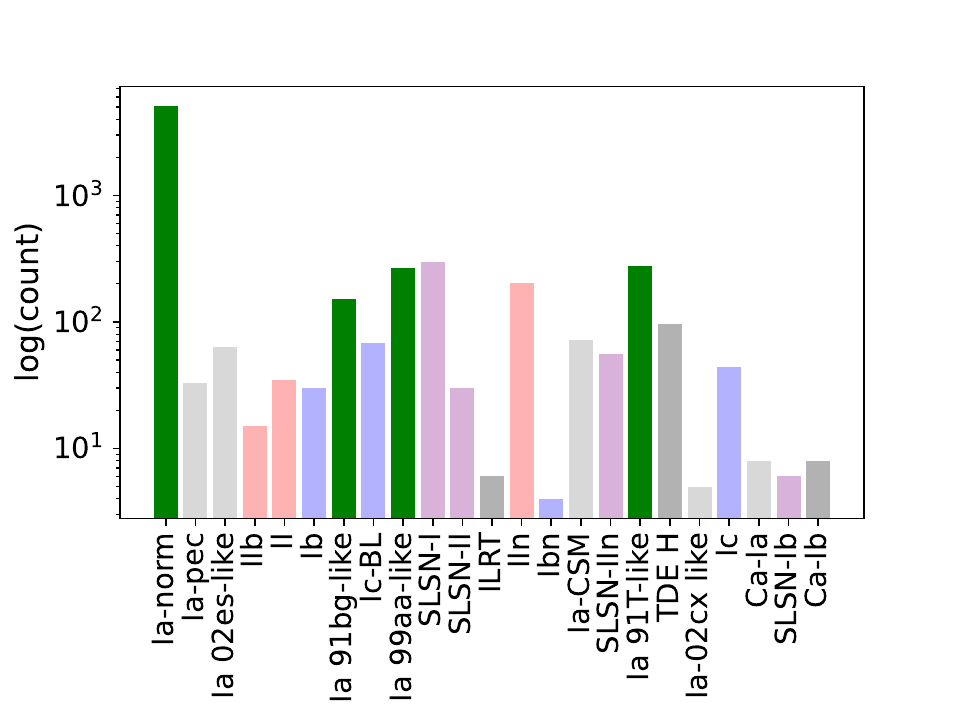}
\caption{\ngsf{} known \(z\)} \label{Ia:NGSF known z}
\end{subfigure}\hspace*{\fill}
\begin{subfigure}{0.48\textwidth}
\includegraphics[width=\linewidth]{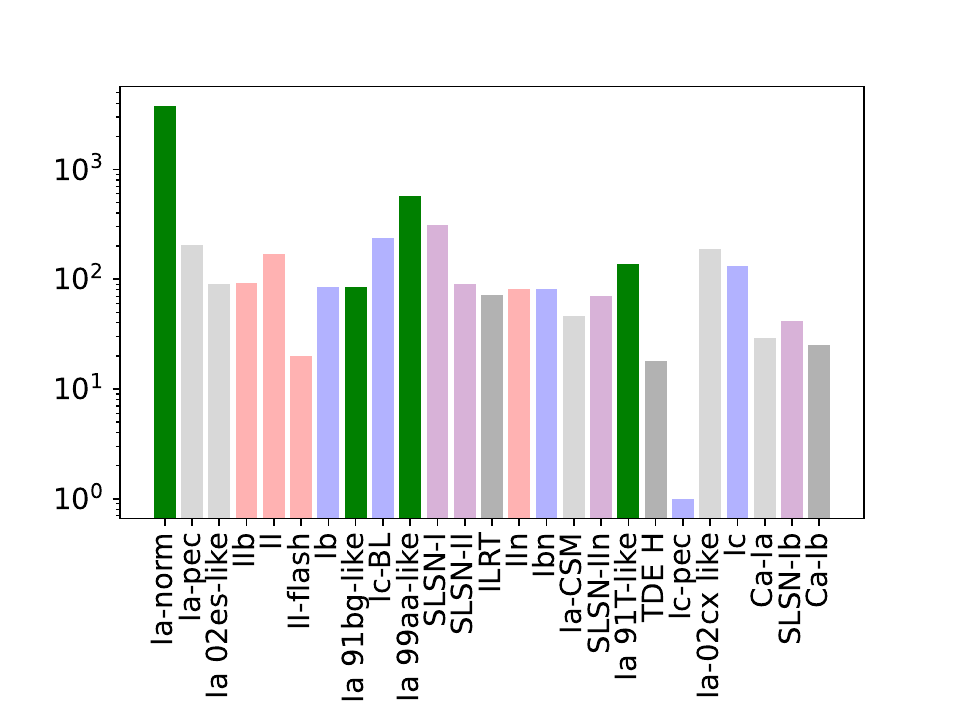}
\caption{\ngsf{} unknown \(z\)} \label{Ia:NGSF unknown z}
\end{subfigure}

\begin{subfigure}{0.48\textwidth}
\includegraphics[width=\linewidth]{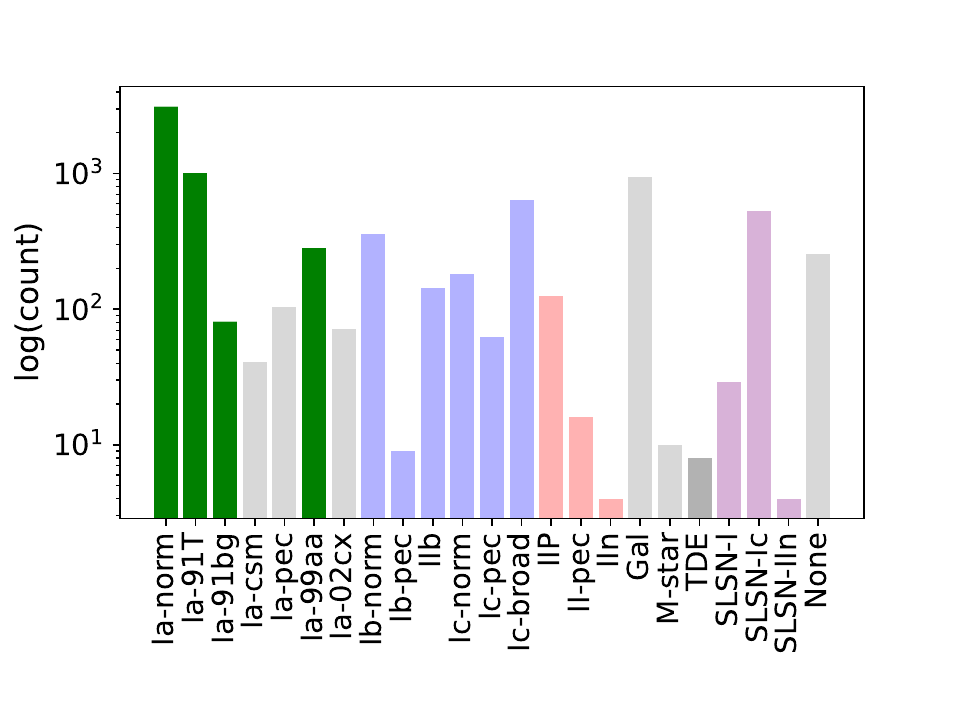}
\caption{\snid{} Known \(z\)} \label{Ia:SNID known z}
\end{subfigure}\hspace*{\fill}
\begin{subfigure}{0.48\textwidth}
\includegraphics[width=\linewidth]{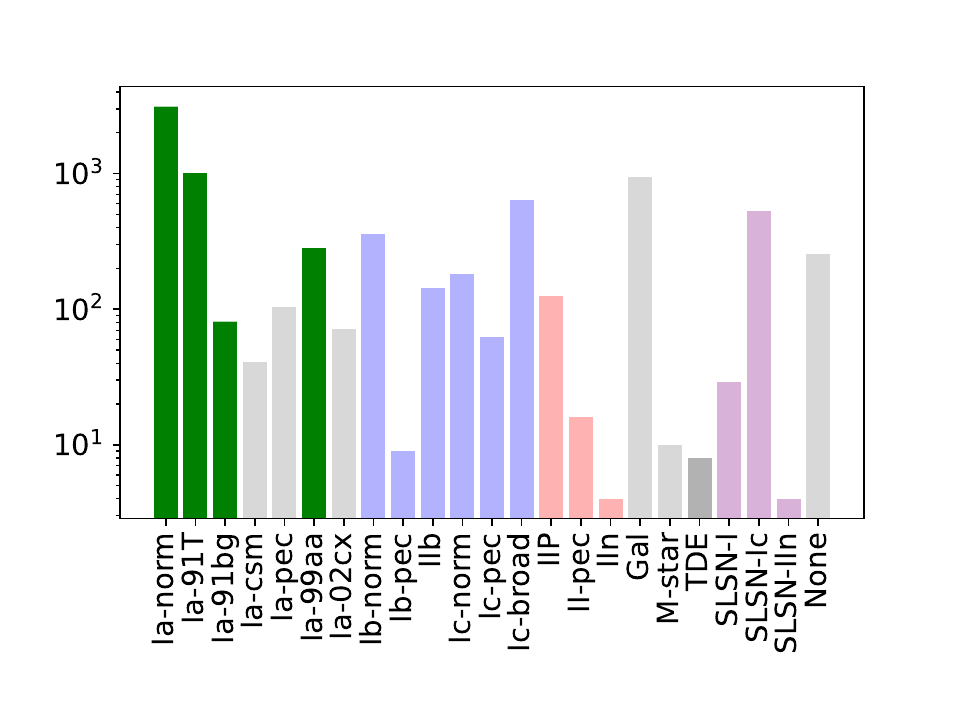}
\caption{\snid{} unknown \(z\)} \label{Ia: SNID unknown z}
\end{subfigure}
\caption{Graphical representation of how SN Ia input spectra are being classified by each classifier with (left column) and without (right column) redshift priors. The subclass of the best-fitting templates is assumed as the subclass of the output. Each histogram lists only the subclasses with at least one output classification. SN Ia subclasses are green, Ibc are blue, II are red, SLSNe are purple, non-SNe are black and `other' classes (Ia-pec, non-transients) are gray. The shift from Ia-norm to other SNe Ia subclasses when redshift priors are removed can be seen.}
\label{Ia fits}
\end{figure*}

Fig. \ref{Ia fits} shows how the SN Ia input spectra are being fit by each classifier, based on the subclass of the best-fitting template. In each case the green bars indicate the good SN Ia classification bins. Non-SN Ia bars of various colours indicate all of the misclassifications. In all three classifiers we investigate we see the same effects of moving from using redshift priors to not. 

There is a shift in successfully classified SNe Ia from the Ia-norm class into other SN Ia and SN Ia-pec subclasses. Additionally the number of SNe Ia incorrectly classified as non-SNe Ia can be seen in the non-green bars universally increasing in height. Boths of these effects serve to diminish the SN Ia classification rate without redshifts.

Of note are the tendency of \dash{} to classify transients as SN Ia-csm. This seems to be the result of narrow galaxy emission lines from Sc host templates masquerading as the narrow lines of an ejecta-csm interaction. The inclusion of SN Ia-csm as an acceptable SN Ia class for \dash{} does improve the SN Ia classification rate, but at the cost of contamination rates exceeding 15\(\%\). A similar effect occurs with \snid{}, except that it does seem to prefer to correctly identify them as galaxies with a `Gal' output.

\begin{figure*}
\begin{subfigure}{0.48\textwidth}
\includegraphics[width=\linewidth]{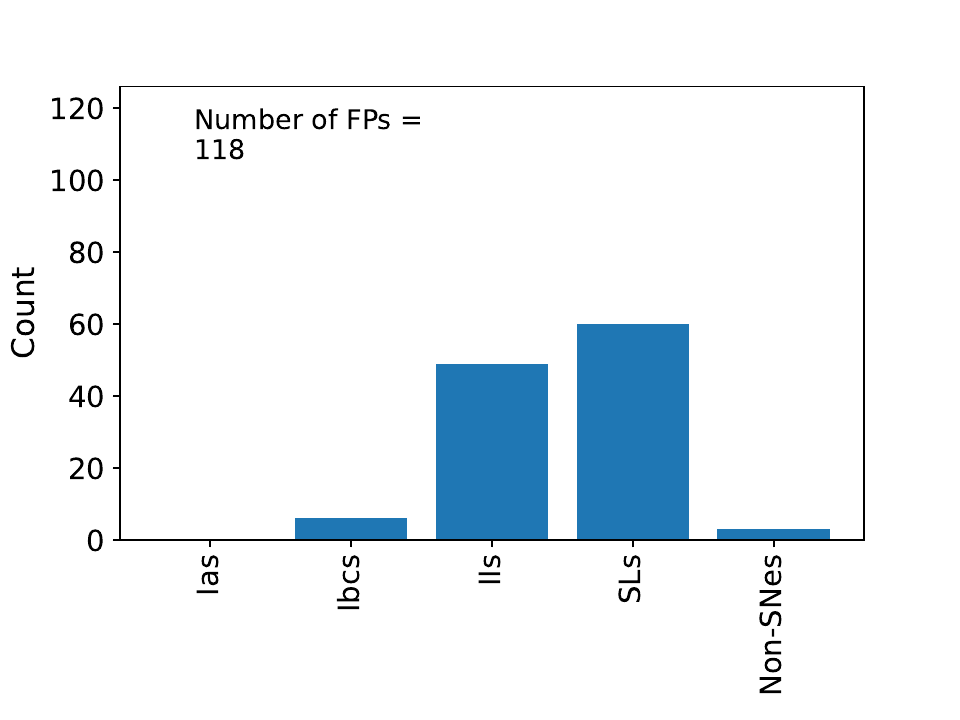}
\caption{\dash{} known \(z\)} \label{contaminant:Dash known z}
\end{subfigure}\hspace*{\fill}
\begin{subfigure}{0.48\textwidth}
\includegraphics[width=\linewidth]{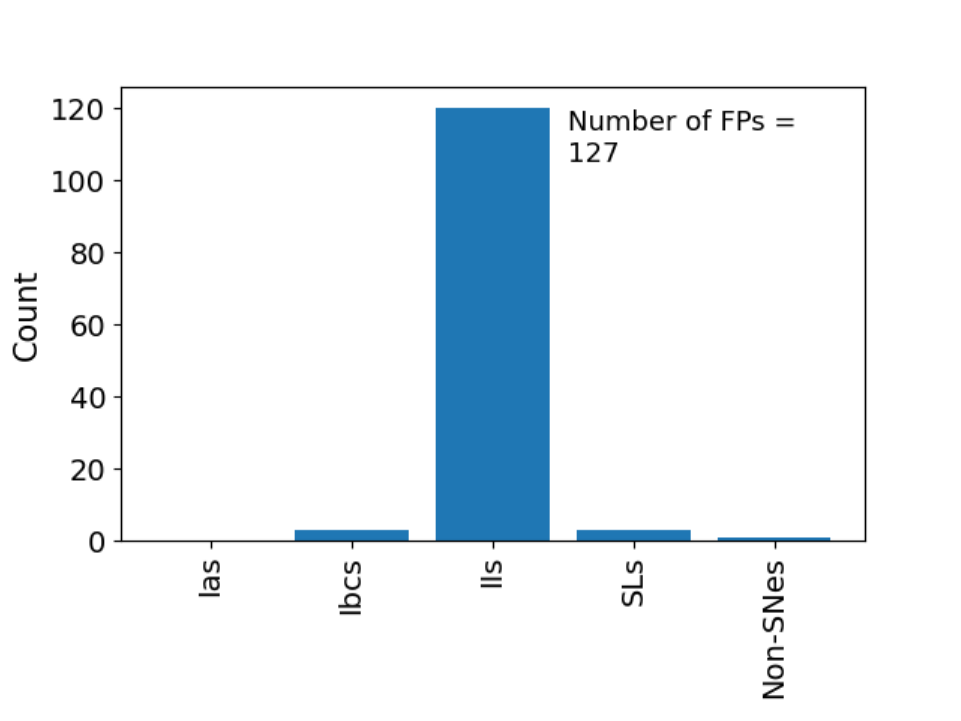}
\caption{\dash{} unknown \(z\)} \label{contaminant:DASH unknown z}
\end{subfigure}

\medskip
\begin{subfigure}{0.48\textwidth}
\includegraphics[width=\linewidth]{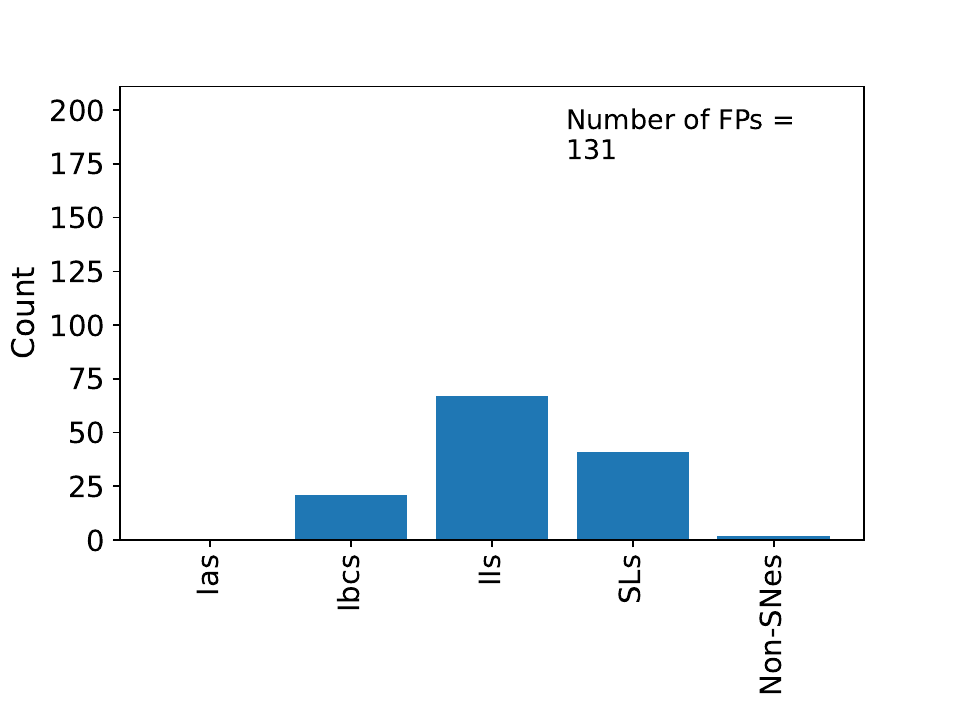}
\caption{\ngsf{} known \(z\)} \label{contaminant:NGSF known z}
\end{subfigure}\hspace*{\fill}
\begin{subfigure}{0.48\textwidth}
\includegraphics[width=\linewidth]{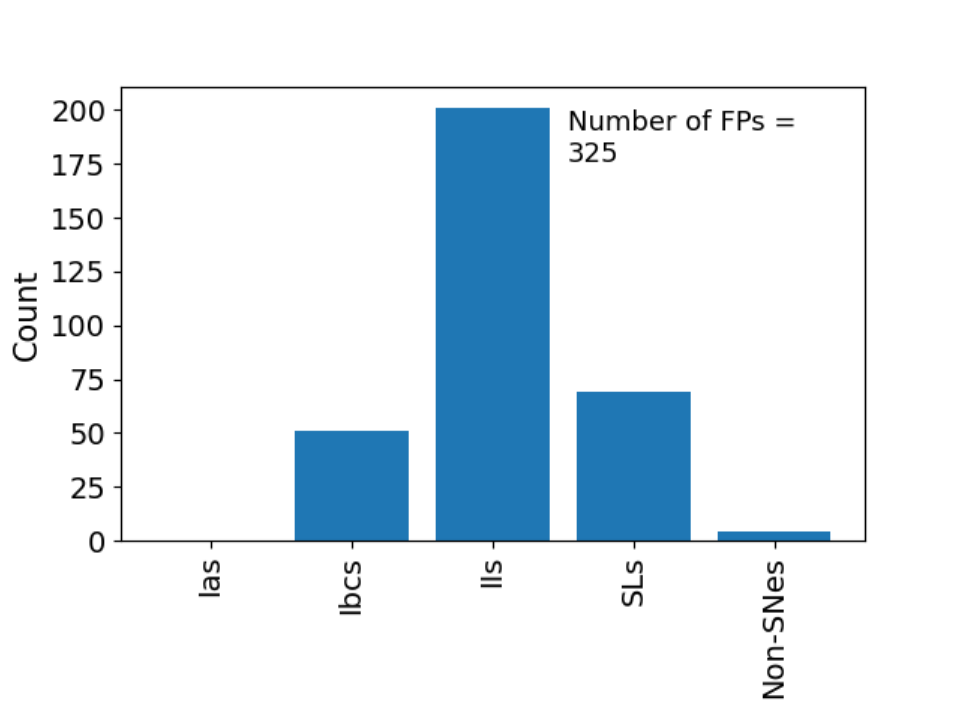}
\caption{\ngsf{} unknown \(z\)} \label{contaminant:NGSF unknown z}
\end{subfigure}

\medskip
\begin{subfigure}{0.48\textwidth}
\includegraphics[width=\linewidth]{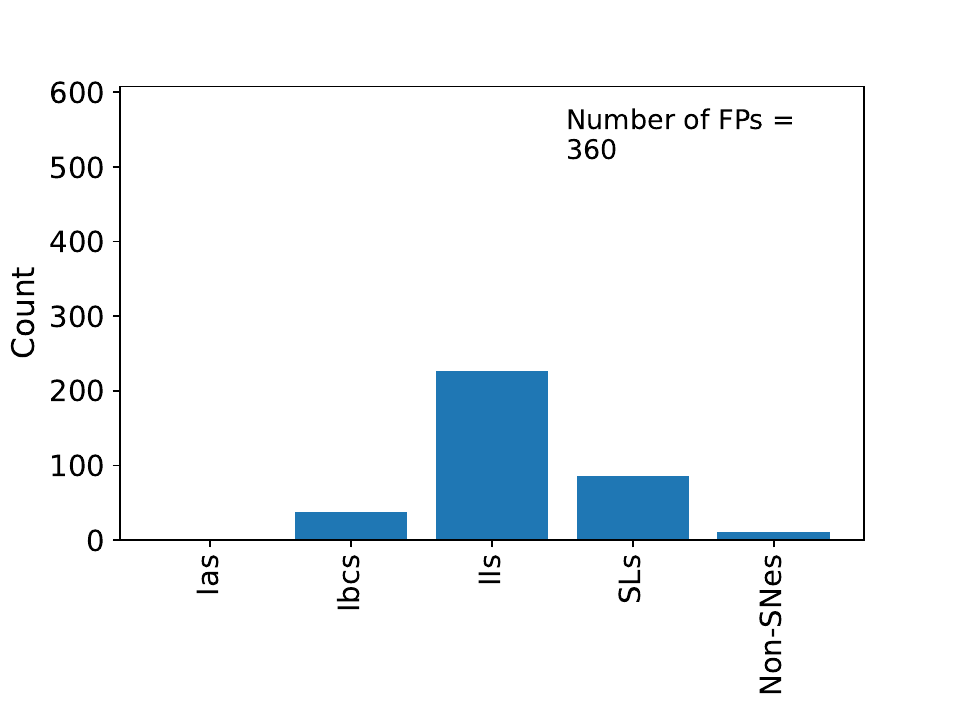}
\caption{\snid{} known \(z\)} \label{contaminant:SNID known z}
\end{subfigure}\hspace*{\fill}
\begin{subfigure}{0.48\textwidth}
\includegraphics[width=\linewidth]{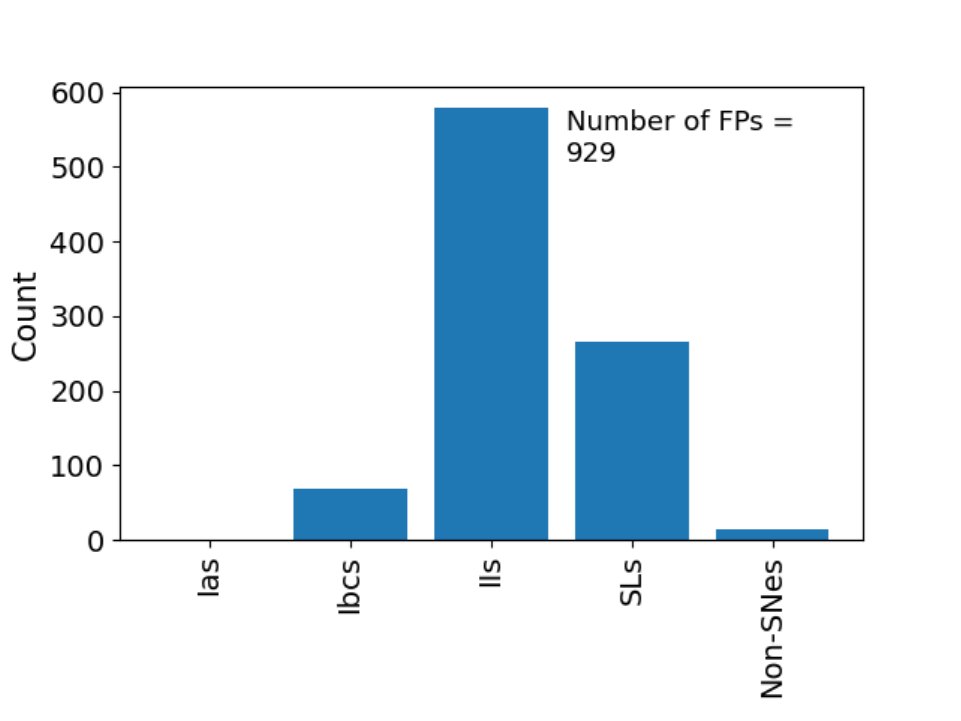}
\caption{\snid{} unknown \(z\)} \label{contaminant: SNID unknown z}
\end{subfigure}
\caption{The distribution of true classifications for objects classified as Ia above the quality threshold to qualify as contaminant results. Input classes are those from the 5-class classification schema. The number of contaminants for each classifier-redshift prior combination are listed on each subplot. The number of FPs increases significantly without redshift priors for \ngsf{} and \snid{}. SLSNe are often over-represented as FPs.}
\label{contaminant fits}
\end{figure*}

Fig. \ref{contaminant fits} shows the origin of the contaminant results for each classifier. We can see immediately that \dash{} suffers as a result of having no ability to classify SLSNe, as they make up the largest fraction of contaminants when redshift priors are known.

When redshift information is removed, \dash{} loses classification performance for all transient classes in both completeness and purity. The fractional decrease in the number of SN Ia and contaminant classification is almost exactly the same, and this results in the purity remaining high (see Table \ref{5class_comp}). The input template classes that produce contaminants is entirely different when redshift priors are removed, now being almost entirely from SNe II. For SLSNe, forcing the classification to high redshifts by using priors resulted in many contaminant Ia classifications. When redshift priors are removed, SLSNe are instead misclassified as other non-SN Ia transients or as SNe Ia-pec. This is a good change from the point of view of SN Ia sample purity.

While we see the contaminant numbers produced by \dash{} maintained when removing redshift knowledge, \ngsf{}  and \snid{} both produce double or more contaminant SN Ia classifications. \ngsf{} and \dash{} both classify predominantly SNe II as contaminant SNe Ia when redshift priors are removed, a significant change from the ratio of classes that produce contaminants with redshift priors. \snid{}'s distribution of contaminants remains almost identical between regimes, although again SNe II are the largest contributor.

Type II SNe are the largest non-SN Ia component of the sample and as expected always dominate the contaminant distribution. In fact, in nearly all cases, the relative number of contaminants originating from the different input non-SN Ia classes at least vaguely mimics their relative abundance in the full sample, slightly shifted by each classifier's ability to classify different classes. Only Fig. \ref{contaminant:DASH unknown z} bucks this trend, producing a large overabundance of SN II contaminant classifications.

\section{Example Classification} \label{App B}

In this appendix we provide some individual classifications as context. We focus on several of the most common types of classification and misclassification. All presented classifications are from \ngsf{} as it is the most prevalent in our suggested classification plan in Section \ref{an example}.

\begin{figure*}
\begin{subfigure}{0.5\textwidth}
\includegraphics[width=\linewidth]{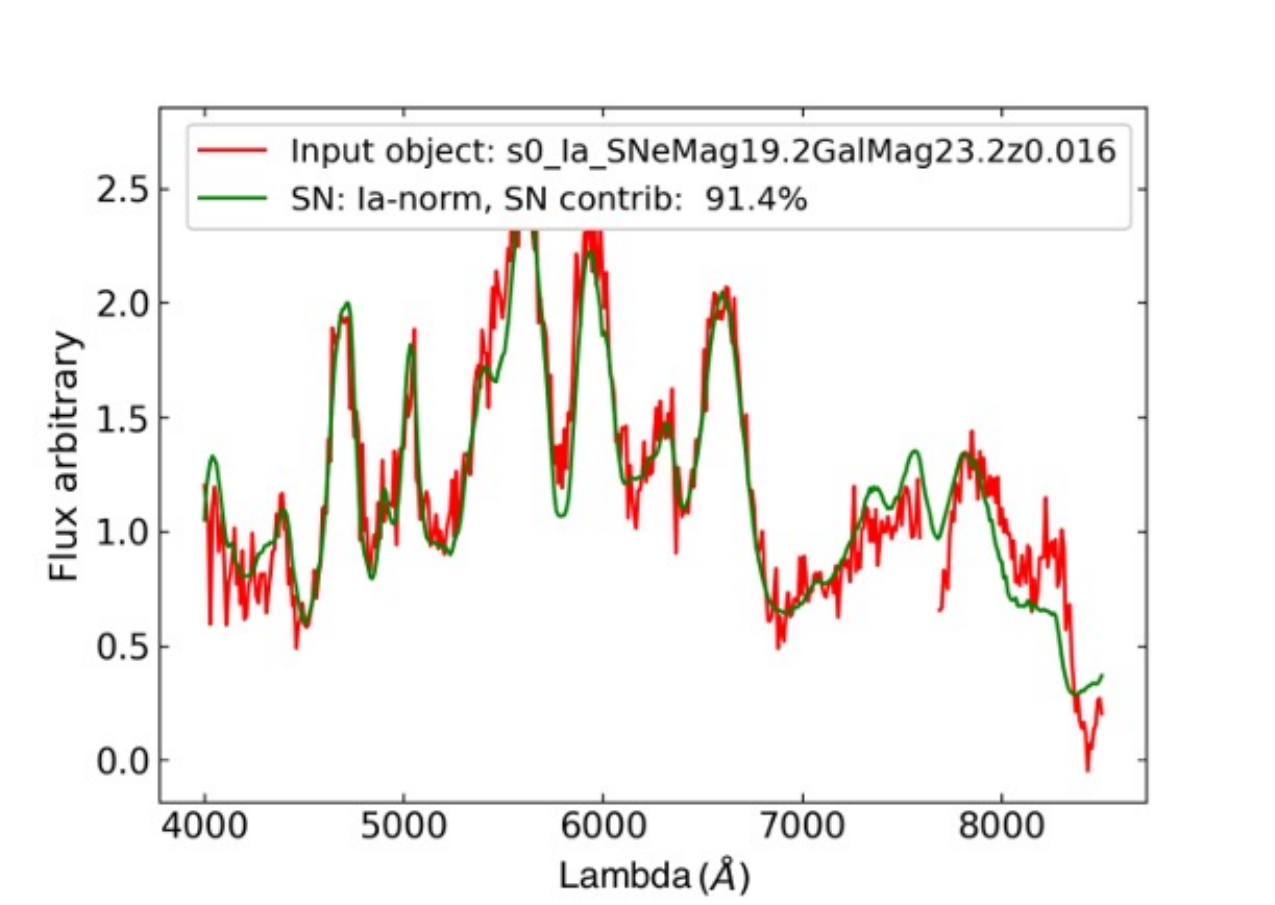}
\end{subfigure}\hspace*{\fill}
\begin{subfigure}{0.5\textwidth}
\includegraphics[width=\linewidth]{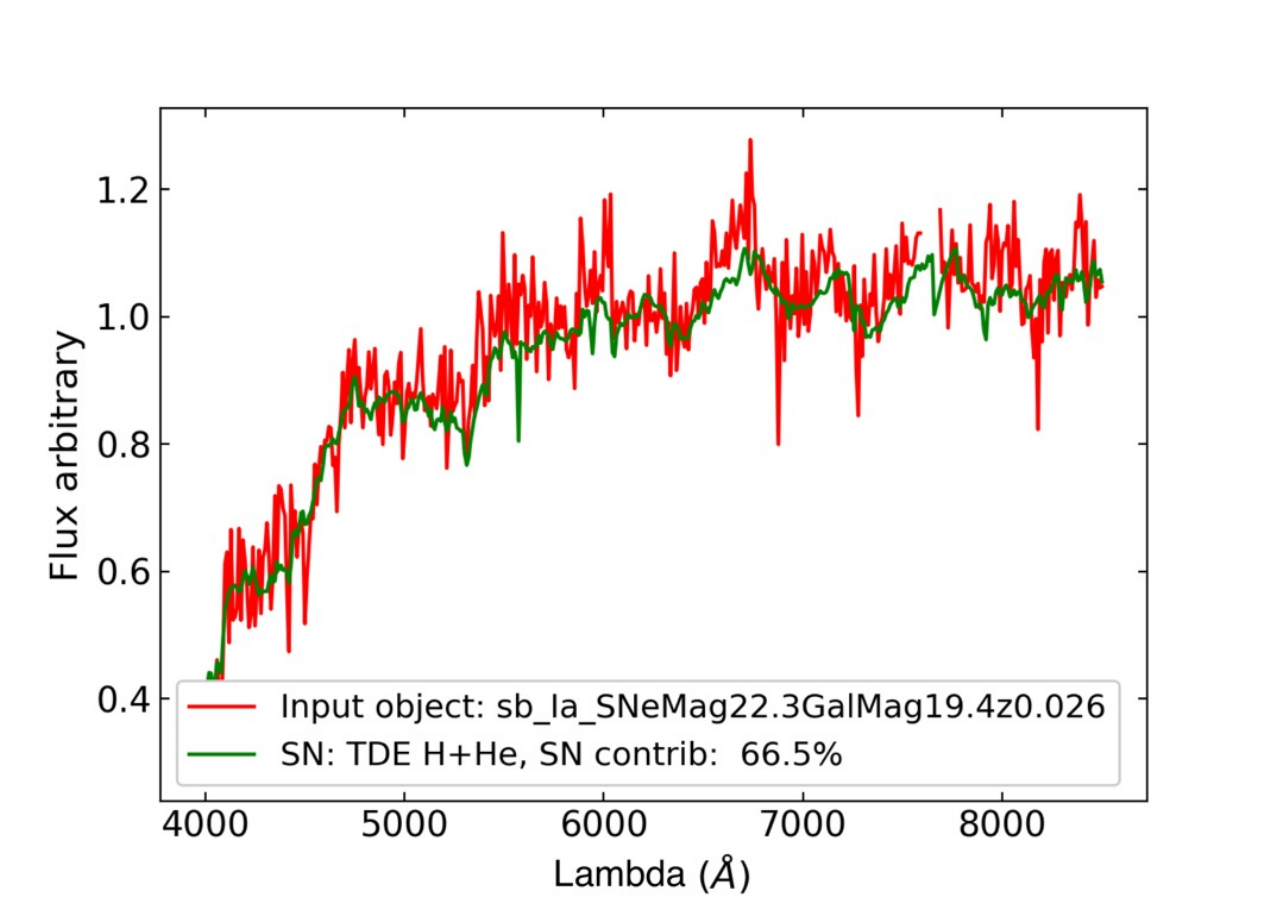}
\end{subfigure}

\medskip
\begin{subfigure}{0.5\textwidth}
\includegraphics[width=\linewidth]{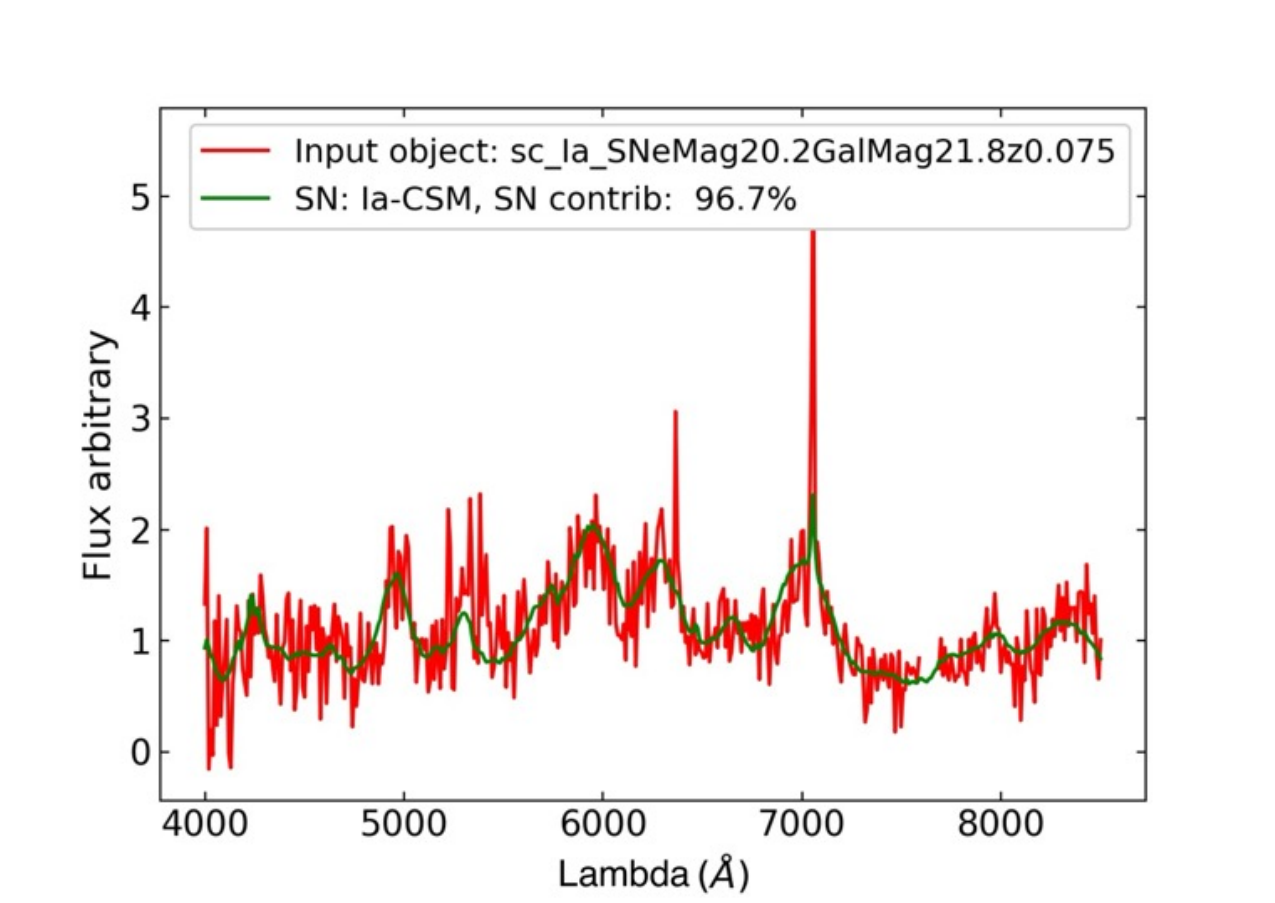}
\end{subfigure}\hspace*{\fill}
\begin{subfigure}{0.5\textwidth}
\includegraphics[width=\linewidth]{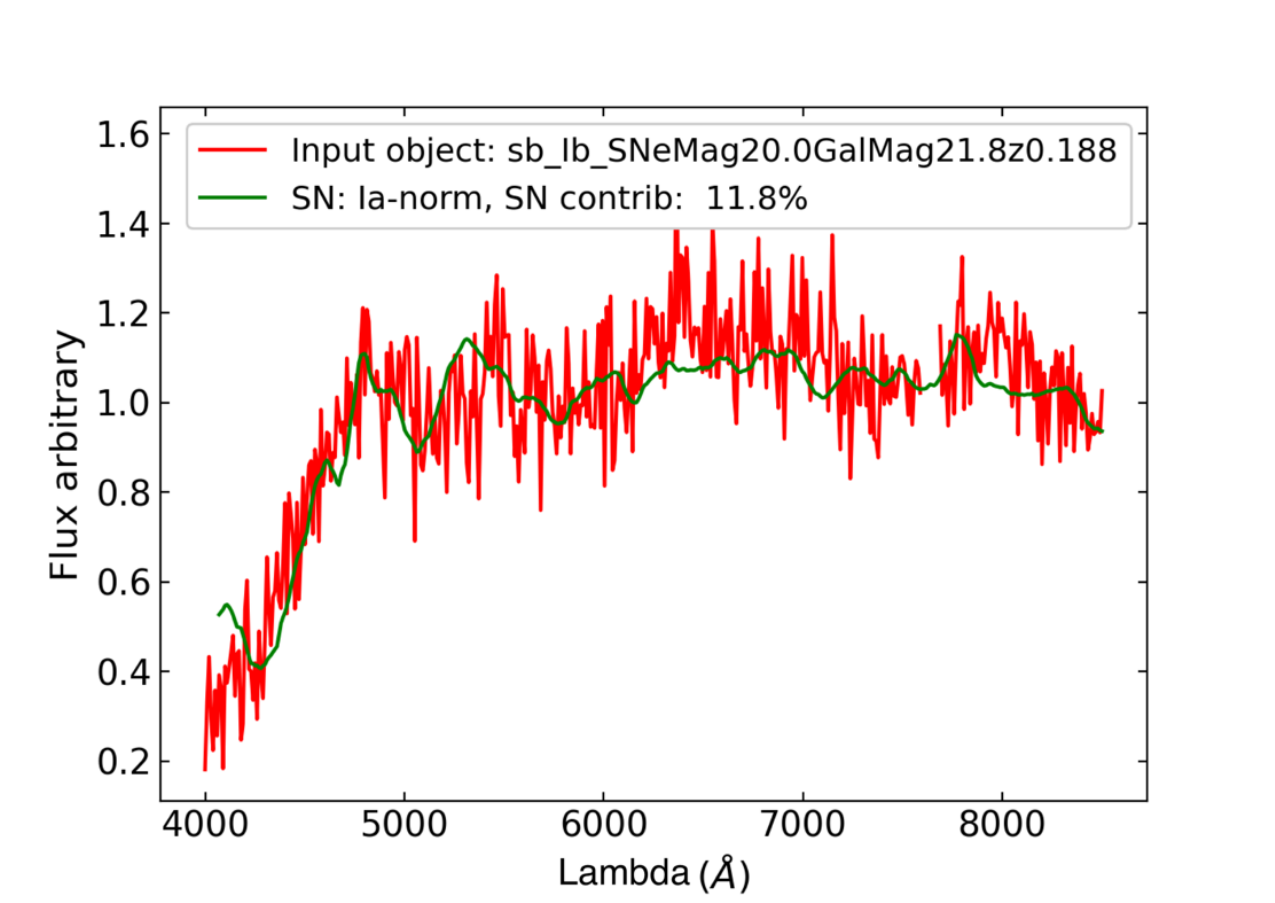}
\end{subfigure}

\caption{Four individual classification results from \ngsf{}. Top left: A good classification of a bright, low contamination SN Ia. Top right: A misclassification of a highly contaminated SN Ia. Bottom left: A misclassification of a bright SN Ia due to narrow galaxy features from its Sc host. Bottom right: An example of a SN Ia false positive where a low contamination SN Ib is misinterpreted as a SN Ia with high contamination. In each case the input is plotted in red with relevant information in the legend. The best-fitting template spectrum is plotted in green and the best-fitting transient class is provided in the legend. The host galaxy fraction of \ngsf{}'s best fitting template is included in the legend with the best fit.}
\label{classifications}
\end{figure*}

Fig. \ref{classifications} shows 4 attempted classifications with \ngsf{}. Fig. \ref{classifications}(a) shows a successful SN Ia classification. We find that noisy spectra, where the transient is faint, or spectra with significant host contamination are often hard to classify as would be expected. This is shown in Fig. \ref{classifications}(b) We also see an overabundance of misclassifications from spectra with the Sc host template. These are often the result of the classifier misinterpreting the strong galaxy emission as narrow features from the transient. This leads to a classifications of SN Ia-csm and other narrow emission transient subclasses like Ibn, IIn etc. This is shown in Fig. \ref{classifications}(c). False positive SN Ia classifications can arise from many effects. Shown in Fig. \ref{classifications}(d) we have a low host contamination SN Ib being misinterpreted as a Ia-norm with significant host contamination. This suggests that there is degeneracy between SN subclass and host contamination levels.

\section{Example Spectra} \label{example spec}

\begin{figure}
    \centering
    \includegraphics[width=\linewidth]{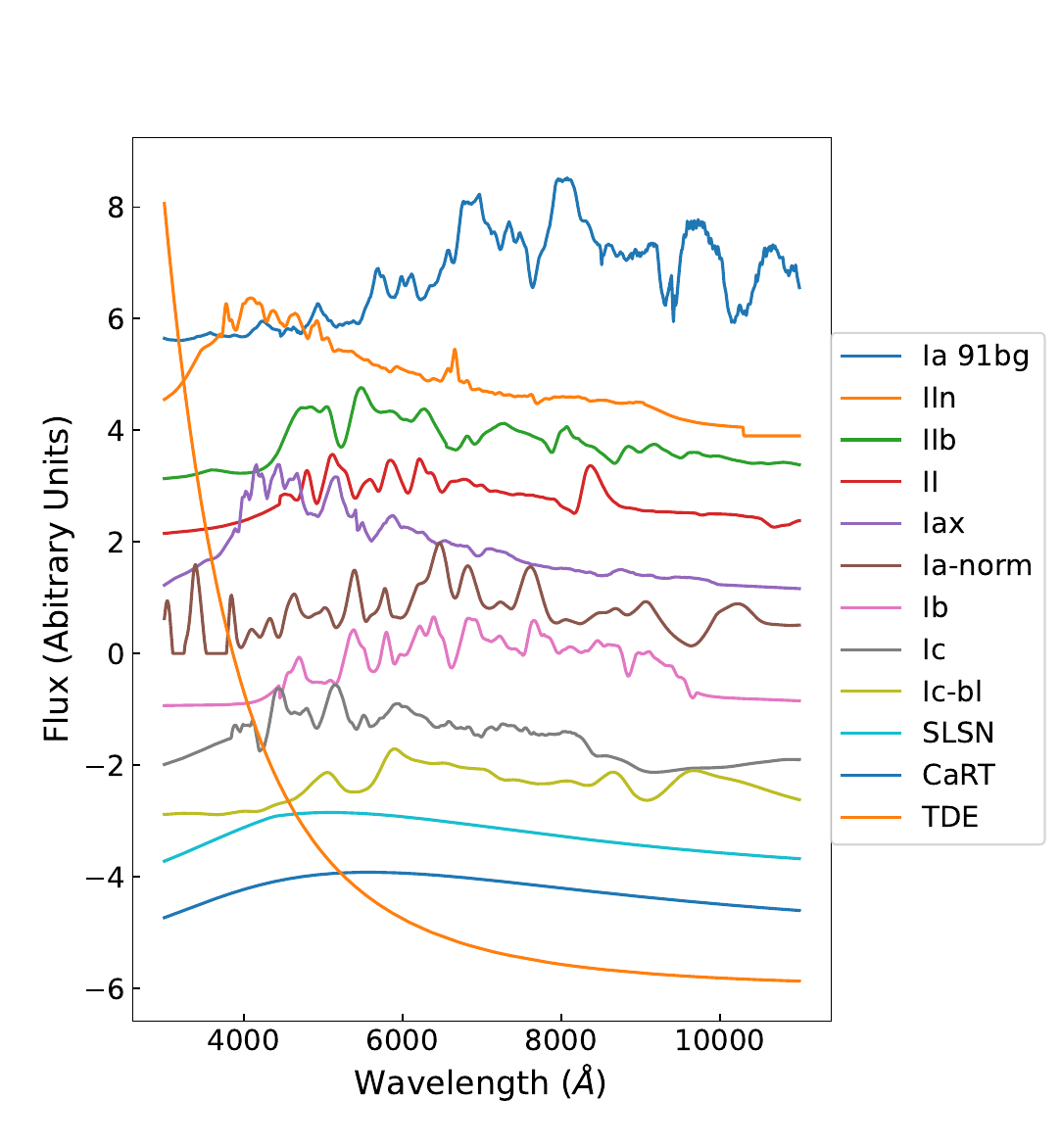}
    \caption{Example spectra for each distinct input transient class. Spectra such as these were used as the starting point to generate the simulated spectra in Section \ref{creating}.}
    \label{ex spec}
\end{figure}

Figure \ref{ex spec} shows an example of each of the twelve types of inputs transient spectra used in our blended spectra simulations. The spectra belong to the transient classes of: Ia-norm, Ia 91bg-like (faint, fast-declining), Iax (faint, progenitor-preserving white dwarf thermonuclear detonations), Ib, Ic, IIb, Ic-BL, II, IIn (all core-collapse SNe), SLSNe (incredibly bright transients), TDEs (star disrupted by black hold tidal forces) and CaRTs (SN Ia-related events, rich in Calcium).

The spectra presented here are arbitrarily scaled and flux-shifted for presentation. No simulated fibre effects or observational noise from the 4MOST ETC has been added. As noted in Section \ref{creating}, the primary purpose of the spectra is for simulating realistic light-curve information for LSST rather than accurately portraying the spectra of a given transient class. Rarely observed transient classes, such as TDEs and CaRTs are essential featureless blue continua, combined with a limited presence in classifier training samples/template banks, is likely partially responsible for their incredibly poor classification results.

\end{document}